\documentclass[lettersize,journal]{IEEEtran}
\usepackage{amsmath,amssymb,amsfonts}
\usepackage[utf8]{inputenc}
\usepackage{algorithmic}
\usepackage{algorithm}
\usepackage{array}
\usepackage[caption=false,font=scriptsize,labelfont=sf,textfont=sf]{subfig}
\usepackage{textcomp}
\usepackage{stfloats}
\usepackage{url}
\usepackage{verbatim}
\usepackage{graphicx}
\usepackage{cite}
\usepackage{amssymb}
\usepackage{setspace}
\usepackage{lineno} 
\usepackage{cases}
\usepackage{subfloat}
\usepackage{float}  
\usepackage{titlesec}
\usepackage{textcomp}
\usepackage{bm,bbold}
\usepackage{xcolor}
\usepackage{siunitx}
\newcommand\bw{\ensuremath{{\bm w}}}

\newcommand\bW{\ensuremath{{\bold W}}}

\newcommand\bq{\ensuremath{{\bm q}}}
\newcommand\bx{\ensuremath{{\bm x}}}
\newcommand\by{\ensuremath{{\bm y}}}

\newcommand\bh{\ensuremath{{\bm h}}}

\newcommand\be{\ensuremath{{\bm e}}}
\newcommand\bz{\ensuremath{{\bm z}}}

\newcommand\bp{\ensuremath{{\bm p}}}

\newcommand\ba{\ensuremath{{\bm a}}}

\newcommand\bb{\ensuremath{{\bm b}}}
\newcommand\bg{\ensuremath{{\bm g}}}

\newcommand\bv{\ensuremath{{\bm v}}}

\newcommand\bV{\ensuremath{{\bold V}}}

\newcommand\bs{\ensuremath{{\bm s}}}

\newcommand\bE{\ensuremath{{\bold E}}}

\newcommand{\bI}{{\bold I}}

\begin{document}

\title{ Hierarchical Sub-Array Beam Training for Flexible Intelligent Metasurface-Enabled Hybrid Near-Far-Field Multiuser Communications
}

\author{Qian Zhang,~\IEEEmembership{Member,~IEEE,}
		Yufei Zhao,~\IEEEmembership{Member,~IEEE,}
		Jiancheng An,~\IEEEmembership{Senior Member,~IEEE,}\\
		Zheng Dong,~\IEEEmembership{Member,~IEEE,}
		Yong Liang Guan,~\IEEEmembership{Senior Member,~IEEE,}
		Ju Liu,~\IEEEmembership{Senior Member,~IEEE,} \\
		and 
		Chau Yuen,~\IEEEmembership{Fellow,~IEEE}
\thanks{
}
\thanks{Qian Zhang is with the School of Computer and Communication Engineering and Hebei Key Laboratory of Marine Perception Network and Data Processing,, Northeastern University at Qinhuangdao, Qinhuangdao 066004, China. He was also with the School of Information Science and Engineering, Shandong University, Qingdao 266237, China (e-mail: zhangqian@neuq.edu.cn).}
\thanks{Zheng Dong and Ju Liu are with School of Information Science and Engineering, Shandong University, Qingdao 266237, China (e-mail: \{zhengdong,juliu\}@sdu.edu.cn).}
\thanks{Yufei Zhao, Yong Liang Guan, and Chau Yuen are with School of Electrical and Electronics Engineering, Nanyang Technological University, Singapore 639798 (e-mail: yufei.zhao@ntu.edu.sg; EYLGuan@ntu.edu.sg; chau.yuen@ntu.edu.sg).}
\thanks{Jiancheng An is with the School of Electronic Science and Engineering, University of Electronic Science and Technology of China (UESTC), Chengdu, 611731, China (e-mail: jiancheng.an@uestc.edu.cn).}
}

\markboth{}%
{Shell \MakeLowercase{\textit{et al.}}: A Sample Article Using IEEEtran.cls for IEEE Journals}


\maketitle

\begin{abstract}
Low-resolution discrete phase shifters (PSs) are attractive for extremely large-scale arrays, but phase quantization distorts near-field (NF) beam patterns and degrades beam training and multiuser transmission. Moreover, the distinct codebook structures required by NF and far-field (FF) users complicate hybrid NF-FF beam training and codeword switching.
To address these challenges, we introduce flexible intelligent metasurface (FIM) surface shaping as an additional geometric beamforming degree of freedom. We establish the hybrid NF-FF channel and beam pattern models, characterize the grating lobes induced by discrete PSs, and develop a phase-quantization compensation method. A deformation-range analysis shows that wavelength-scale shaping can achieve beam patterns comparable to those obtained with continuous PSs. We further propose a hierarchical sub-array beam training (HSABT) scheme that progressively doubles the activated central aperture according to the beam-width and beam-depth characteristics. A beam-intersection-based sampling strategy improves the angle-distance resolution, while a measurement-driven adaptive NF-FF beam-switching mechanism selects the final codeword without relying on the Rayleigh distance or prior user-location information. HSABT is also extended to multi-path channels through adaptive peak detection and inter-level energy-consistency-based candidate tracking. For high-gain multiuser transmission, we develop an alternating-optimization-based design and a complementary non-iterative key observation method, and analyze their convergence, computational complexity, and steady-point properties.
Simulation results demonstrate continuous-phase-comparable beam patterns, reduced training overhead, more accurate main-path acquisition than exhaustive polar-codebook training, and higher multiuser sum rates than conventional rigid-array benchmarks.

\end{abstract}


\section{Introduction}
Extremely large-scale arrays (XL-arrays) have emerged as a promising technology for future wireless communication systems due to their potential to significantly enhance spectral efficiency and spatial resolution~\cite{Lu2024XL_MIMO,Chen2025XL_MIMO,Cui2023NF_MIMO,Chen2024THBF,Tarboush2024Cross_Field}. By deploying a massive number of antenna elements over an electrically large aperture, XL-arrays can exploit fine-grained spatial degrees of freedom for high-resolution beamforming, user separation, and environmental sensing. However, as the array aperture expands, the conventional plane-wave model becomes inaccurate, and a spherical-wave model is required to capture the non-negligible wavefront curvature~\cite{An2024NF_Comm,Yang2024XL_RIS,Tarboush2024Cross_Field}. This propagation transition extends communications from the conventional far-field (FF) regime to the near-field (NF) region, where distance-dependent phase variations and spatial non-stationarity must be explicitly considered. The NF-FF boundary is commonly characterized by the Rayleigh distance~\cite{Liu2023NF_Review,Cui2023NF_MIMO}.

However, the expansion of the NF region introduces new challenges for channel estimation and beamforming design.
The channel dimensionality increases with both the number of antennas and the considered distance range, making full channel reconstruction increasingly costly for XL-arrays. To address this issue, codebook-based beam training (BT) has emerged as an efficient approach for acquiring channel state information (CSI)~\cite{Lu2024HBT,Lim2020EBT,Zheng2025NFWBT,Xiao2016H_Codebook}.
Instead of estimating the complete channel, BT identifies dominant propagation paths and focal points, thereby significantly reducing pilot and feedback overhead. With its compatibility with hybrid beamforming architectures and robustness to calibration errors and NF spatial non-stationarity, BT has been widely adopted and standardized in IEEE 802.15.3c and IEEE 802.11ad~\cite{Perahia2010802,Xiao2016H_Codebook,Wang2009Beam}.

Nevertheless, NF channels jointly depend on the angular and distance domains, rendering conventional angular-domain DFT codebooks inadequate. To address this issue, Cui {\it et al.}~\cite{Cui2022PolarCodebook} developed a polar-domain codebook that jointly samples angle and distance, enabling accurate and noise-robust exhaustive NF-BT. However, its training overhead grows rapidly with the array size. To reduce this overhead, several hierarchical and two-stage schemes have been proposed~\cite{Zhang2025XL_RIS,Zhang2022TPBT,Chen2024BT}. Specifically, the multi-resolution codebook in~\cite{Zhang2025XL_RIS} progressively refines the search using wide-to-narrow beams, albeit with reduced noise robustness. The method in~\cite{Zhang2022TPBT} first identifies candidate angles and then performs distance-domain refinement, while~\cite{Chen2024BT} employs sub-arrays for efficient training in hybrid NF and FF environments.

In practice, XL-arrays typically employ low-resolution PSs to reduce the hardware cost, power consumption, and control complexity associated with a massive number of radiating elements~\cite{Lu2024HBT,Zhang2025NF_DPS}. 
However, phase quantization prevents accurate matching of NF steering vectors, resulting in residual phase errors and strong grating lobes. 
The Fourier series expansion-based analysis in~\cite{Zhang2025NF_DPS} showed that the grating-lobe gain increases as the phase resolution decreases. 
This degradation is particularly severe in hierarchical sub-array BT, where quantization errors may propagate across successive training stages and impair the acquired CSI. 
Thus, discrete phase shifters (PSs) provide a practical hardware-accuracy trade-off, but may substantially degrade subsequent multiuser performance.

Employing higher-resolution PSs is a direct solution, but it increases the electronic control complexity while retaining a fixed array geometry. 
Similarly, conventional reconfigurable intelligent surface (RIS)~\cite{Zhang2023Robust_RIS,Zhang2024Practical_RIS} and rigid XL-arrays can reconfigure their electromagnetic coefficients but cannot alter their element locations or propagation paths. 
Hence, phase-resolution enhancement and geometric reconfiguration constitute two distinct design dimensions: the former improves electronic phase accuracy over a fixed aperture, whereas the latter modifies the propagation distances and the resulting NF array manifold. 
Recent fluid antennas~\cite{Zhang2025FA_ISAC,Li2026FA_SecureISAC}, movable antennas~\cite{Wang2026MA_OTFS}, and shape-reconfigurable arrays have further demonstrated the potential of exploiting spatial and geometric DoFs. 
However, geometric reconfiguration also changes the effective aperture and dynamically shifts the NF-FF boundary, making communication design for hybrid NF and FF users substantially more challenging.
Against this background, flexible intelligent metasurfaces (FIMs) with controllable deformation have emerged as a new class of geometry-reconfigurable apertures~\cite{Rao2021NE,Ni2022NC,Wang2021FIM}. 
Unlike rigid XL-arrays and conventional RISs, FIMs introduce the surface geometry as an additional controllable DoF~\cite{Bai2022FIM,An2025FIM_TAP,An2025FIM}. 
For the considered linear FIM, each radiating element can be displaced perpendicular to the original aperture, thereby modifying its propagation distances to the users. 
Since the NF steering phase depends on both the user location and element position, wavelength-scale deformation can generate controllable propagation-phase variations and reshape the range-angle-dependent beam pattern. 
This geometric phase adjustment complements low-resolution electronic phase control and enables compensation for phase-quantization errors without relying exclusively on higher-resolution PSs.

Nevertheless, existing FIM studies mainly focus on received-power enhancement, channel estimation, or capacity optimization under general communication models~\cite{Bai2022FIM,Zhang2026FIM_CRB,An2025FIM}. 
The effects of wavelength-scale deformation on NF beam width, beam depth, grating lobes, and angle-distance coupling under low-resolution phase control remain insufficiently understood. 
Moreover, existing XL-RIS and XL-array beam-training methods generally assume a fixed aperture, while low-resolution beamforming designs do not exploit geometric deformation for phase-error compensation. 
Therefore, it remains challenging to determine whether limited deformation is sufficient for quantization-error compensation, efficiently acquire hybrid NF and FF users under a dynamically varying aperture, and jointly optimize the continuous deformation and discrete analog beamforming variables.

Building on the above motivation, we investigate FIM-enabled hybrid near-far-field BT and multiuser transmission. By jointly exploiting low-resolution electronic phase control and wavelength-scale geometric adaptation, the proposed framework aims to compensate for phase-quantization errors, enable low-overhead user acquisition across the near- and far-field regions, and achieve high-gain multi-beam transmission. The main contributions are summarized as follows.
\begin{itemize}
	\item[$\bullet$] We introduce FIM surface shaping as an additional geometric beamforming DoF for hybrid near-far-field communications. Specifically, we establish the corresponding FIM channel model and characterize its NF and FF beam patterns, based on which the grating-lobe effects induced by discrete phase shifters are analyzed. We then develop a phase-quantization model for the near-field steering vector and design the FIM surface shaping to compensate for the resulting quantization errors. Furthermore, we derive an error-compensation-based deformation-range criterion, demonstrating that wavelength-scale surface shaping is sufficient to achieve beam patterns comparable to those obtained with continuous PSs.
	
	\item[$\bullet$] We develop a hierarchical sub-array beam training (HSABT) framework for hybrid near-far-field and multi-path channel acquisition. HSABT progressively doubles the activated central aperture based on the beam-width and beam-depth characteristics and propose beam-intersection-based sampling method to improve the angle-distance resolution. Importantly, we propose a measurement-driven adaptive near-far-field beam-switching mechanism that selects the final NF or FF beam according to the measured beamforming gains, without relying on the Rayleigh distance or prior knowledge of the user location. For multi-path channels, cubic-spline-based peak detection and an inter-level energy-consistency criterion are introduced to adaptively determine and track multiple candidate paths without prior knowledge of the path number. The proposed scheme suppresses spurious candidates and achieves higher main-path acquisition accuracy than exhaustive polar-domain codebook training under the considered settings, while significantly reducing the training overhead.
		
	\item[$\bullet$] We propose two complementary beamforming design methods for high-gain multiuser transmission. The first is an alternating-optimization-based method that sequentially updates the desired beam gains, discrete analog beamformers, and bounded FIM surface shaping variables. The discrete phase constraints are handled through a proximal-distance update, while the surface shaping subproblem is solved using the proposed $\rm tanh$ reparameterization method. We establish the convergence of the objective sequence and characterize the resulting solution as a block-wise steady point. The second method, termed the key observation method (KOM), transforms a simplified surface shape design into independent per-element quartic-polynomial problems and obtains the global optimum of the resulting separable surrogate by evaluating the feasible stationary points and deformation boundaries. As a non-iterative alternative, KOM substantially reduces the computational burden while maintaining competitive multiuser transmission performance. The convergence behavior, computational complexity, and numerical stability of the proposed methods are further analyzed and validated through simulations.

\end{itemize}

{\it{Notation:}} 
$\mathbb{C}^{M \times N}$ denotes the space of $M \times N$ complex matrices.
$\mathbb{Z}$ denotes the set of integers.
$\mathbb{Z}^+$ denotes a set of positive integers.
$[\ba]_n$ represents the $n$-th entry of the vector $\ba$.
$\bx \sim {\cal{CN}}\left( \bm 0, \bm\Gamma \right)$ represents that $\bx$ follows a circularly symmetric complex Gaussian (CSCG) distribution with mean vector $\bm 0$ and covariance matrix $\bm\Gamma$.
$||\cdot||_2$ and $|\cdot|$ denote the Euclidean norm and absolute value, respectively.
$(\cdot)^{*}$, $(\cdot)^{\rm T}$, $(\cdot)^{\rm H}$, and $(\cdot)^{-1}$ denote the complex conjugate, transpose, and Hermitian transpose, matrix inverse, respectively.
$\odot$ and $\otimes$ denote the Hadamard product and the Kronecker product, respectively.
$\bold{I}_K$ is the $K \times K$ identity matrix.
$\Re\{\cdot\}$ and $\Im\{\cdot\}$ denote the real and imaginary parts, respectively.
${\rm mod}(\cdot,\cdot)$ denotes the modulo operation and $\lfloor \cdot \rfloor$ denotes the floor operation.

\section{System Model}
\begin{figure*}[t]	
	\centering \includegraphics[width=\linewidth]{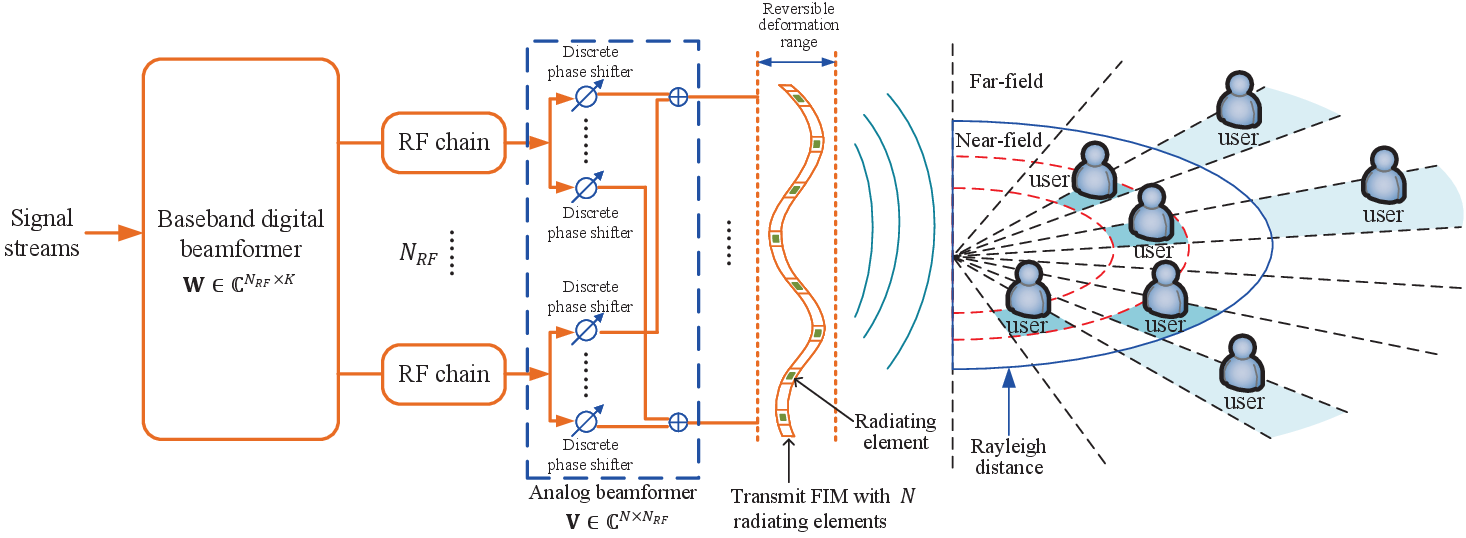}
	\caption{The system model for FIM enabled hybrid NF and FF multiuser communications.}
	\label{fig:System_Model}
\end{figure*}

\begin{figure}[t]
	\centering
	\vspace{-0.25cm}
	\subfloat[Channel model]{\includegraphics[width=0.48\columnwidth]{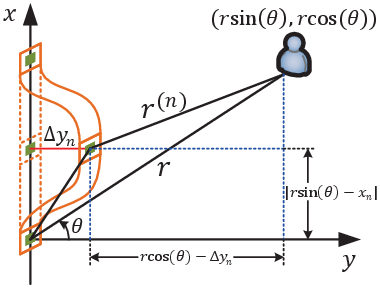}%
		\label{Channel_Model_sub}}
	\hfil
	\subfloat[Equivalent coordinates]{\includegraphics[width=0.32\columnwidth]{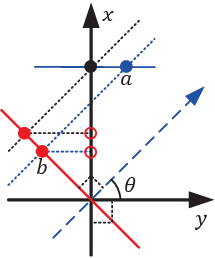}%
		\label{FIM_Rayleigh_distance}}
	\caption{The channel model and equivalent coordinates for FIM enabled NF/FF.}
	\label{fig:Channel_Model}
\end{figure}

We consider a XL-FIM-enabled hybrid NF and FF multiuser downlink system, as shown in Fig.~\ref{fig:System_Model}.
In such a system, the BS equipped with a linear extremely large-scale FIM (XL-FIM) having $N$-radiating elements serves $K$ single-antenna users\footnote{
		The considered FIM architecture is particularly suitable for electrically large millimeter-wave or terahertz access points serving NF and FF users in low-mobility scenarios, such as indoor hotspots, industrial networks, and fixed wireless access. In such settings, the surface shape can be updated on a slow geometry or large-scale-channel timescale, while the phase states and digital beamformers are reconfigured more frequently. Thus, FIM is best viewed as a complementary electronic-geometric architecture, whereas rigid high-resolution arrays may be preferable in high-mobility or ultra-fast reconfiguration scenarios.
}.
Typically, as the array size increases, the Rayleigh distance correspondingly increases, causing an expansion of NF region.
Specifically, the Rayleigh distance $D_{rd}$ can be derived by $D_{rd} = 2 D_{ap}^2 / \lambda$, where $\lambda$ and $D_{ap}$ denote the wavelength and array aperture, respectively.
As shown in Fig.~\ref{fig:Channel_Model}$\,$(a), the coordinate of the $n$-th FIM radiating element is given by $(x_n, \Delta y_n)$.
Based on the geometric relationship illustrated in Fig.~\ref{fig:Channel_Model}$\,$(b), the coordinates of the point corresponding to the projection of $(x_n, \Delta y_n)$ onto the line passing through the origin in the direction with the angle of departure (AoD) $\theta$ are calculated as $(x_n^{\rm p},y_n^{\rm p})$, where $x_n^{\rm p} = x_n\cos^2\theta - \Delta y_n \sin\theta \cos\theta$ and $y_n^{\rm p} = x_n\cos\theta \sin\theta - \Delta y_n \sin^2 \theta$.
The equivalent array aperture in the direction of $\theta$ can be obtained as 
$$
\max\limits_{m,n\in\{1\leq m \leq N, 1\leq n \leq N, m,n\in\mathbb{Z}\}} ~ \sqrt{(x_m^{\rm p} - x_n^{\rm p})^2 + (y_m^{\rm p} - y_n^{\rm p})^2}.
$$
Then, in the XL-FIM system, the Rayleigh distance varies with the deformation of the FIM. 
As a result, some users may switch between the NF and FF regions as the FIM shape changes. 
Therefore, cross-field communication of users is crucial~\cite{Tarboush2024Cross_Field}.
The BT method proposed in this paper enables adaptive switching between NF and FF codewords and operates independently of the Rayleigh distance. 
Therefore, even when the Rayleigh distance changes due to FIM surface shaping, the proposed method remains effective. 
This property will be discussed in detail in Section IV.

\subsection{Signal Model}
To reduce hardware cost and energy consumption, we are considering implementing a hybrid beamforming architecture in the XL-FIM-enabled communication system, which consists of $N_{RF}$ RF chains~\cite{Ahmed2018HB,Zhang2025NF_DPS}, as shown in Fig.~\ref{fig:System_Model}.
Let $\bW \in \mathbb{C}^{N_{RF}\times K}$ and $\bV \in \mathbb{C}^{N\times N_{RF}}$ denote the digital beamformer and the analog beamformer, respectively, which are employed for multiuser interference management and spatial beam focusing.
In addition, we concentrate on discrete PSs, with particular emphasis on the 1-bit PS implementation.
Consequently, the phase of each element in the analog beamformer can only be chosen from a finite set, i.e., $\angle\left( [\bV]_{i,k} \right) \in {\cal C}_b$.
Specifically, for a $b$-bit PS, the finite phase set is denoted as ${\cal C}_b \triangleq \left\{ 0, \frac{2\pi}{2^b},\dots, \frac{(2^b-1)2\pi}{2^b} \right\}$.
Then, the transmit signal of the BS is $\bx = \bV\bW \bs$, where $\bs\in \mathbb{C}^{K\times 1}$ contains $K$ unit-power data streams intended for the $K$ users and $\mathbb{E}[ \bs\bs^{\rm H} ] = \bI_K$.
The received signal at user $k$ is given by \vspace{-0.15cm}
\begin{equation} \label{received_signal}
	\begin{split}
		y_k =& \underbrace{ \bh_k^{\rm H} \bV \bw_{k} s_{k} }_{ {\rm{desired}}\, {\rm{signal}} }  + \underbrace{  \sum_{i =1,i\neq k}^{K} \bh_k^{\rm H} \bV \bw_{i} s_{i} }_{  {\rm{interference}} \, {\rm{from}} \, {\rm{other}} \, {\rm{users}} } + \underbrace{n_k}_{ {\rm{noise}} },
	\end{split}
\end{equation}
where $\bW = \left[\bw_1,\bw_2,\dots, \bw_K \right]$, $n_k \sim {\cal CN}\left( 0, \sigma_k^2 \right)$, and $\bh_k$ denotes the channel between the XL-FIM and the user $k$.

Therefore, the signal-to-interference-plus-noise ratio (SINR) of user $k$ is given by
\begin{equation}
	\begin{split}
		\gamma_k = \frac{ \bh_k^{\rm H} \bV \bw_k \bw_k^{\rm H} \bV^{\rm H} \bh_k }{\sum_{i=1,i\neq k}^{K}\bh_k^{\rm H} \bV \bw_i \bw_i^{\rm H} \bV^{\rm H} \bh_k + \sigma_k^2 },k=1,\dots,K.
	\end{split}
\end{equation}

{\it Remark 1:} 
The number of RF chains $N_{\rm RF}$ is a fixed hardware parameter and generally does not need to match either the total number of associated users or the number of active users in each transmission slot. Let $K_{\rm tot}$ denote the total number of associated users and $K$ denote the number of simultaneously scheduled active users. 
For multiuser spatial multiplexing, the number of transmitted data streams should satisfy $K_{\rm tot} \leq N_{\rm RF}$. 
When $K_{\rm tot}>N_{\rm RF}$, a conventional user-scheduling mechanism can be employed to select at most $N_{\rm RF}$ users for simultaneous transmission in each slot. 
Therefore, the proposed hybrid beamforming framework accommodates a varying number of active users without changing the fixed RF-chain architecture. In particular, the proposed beamforming design is applicable to the general case $K\leq N_{\rm RF}$. The fully loaded case $K=N_{\rm RF}$ is adopted only in the numerical simulations to evaluate the system performance under maximum RF-chain utilization~\cite{Liu2023NF_Review,Zhang2025NF_DPS,Zhou2024MBT}.

\subsection{Channel Model}
In this paper, we investigate high-frequency communication, such as millimeter-wave (mmWave) and terahertz (THz) ranges, where signal propagation is primarily dominated by the line-of-sight (LOS) path~\cite{Zhang2025XL_RIS,Elayan2020Terahertz,Zhang2025NF_DPS}. 
For analytical clarity, we first consider a linear XL-FIM arranged along the $x$-axis, with its origin located at the center of the FIM, as shown in Fig.~\ref{fig:Channel_Model}.
We then use the uniform spherical wave (USW) model~\cite{Liu2023NF_Review} to characterize the channel $\bh_k$, which is given by 
\footnote{
		The considered framework can be extended to a planar FIM and three-dimensional user locations. Specifically, the position of the $(m,n)$-th element can be represented by $\bp_{m,n}=[x_m,\Delta y_{m,n},z_{n}]^{\rm T}$, where $\Delta y_{m,n}$ contains the morphing displacement, while the three-dimensional user location is $\bm\psi_k=r_k[\cos\theta_k\cos\upsilon_k,\cos\theta_k\sin\upsilon_k,\sin\theta_k]^{\rm T}$, with $r_k$, $\upsilon_k$, and $\theta_k$ denoting the distance, azimuth angle, and elevation angle, respectively. The corresponding spherical-wave propagation distance is $d_{k,m,n}=\|\bm\psi_k-\bp_{m,n}\|_2$.
		Accordingly, the Rayleigh distance can be evaluated using the maximum planar aperture, while a distance-azimuth-elevation codebook can be constructed by uniformly sampling the two angular direction cosines and the reciprocal-distance domain. 
		The proposed framework can be generalized by replacing the distance-angle codebook with a distance-azimuth-elevation codebook, while the element-wise FIM compensation principle remains unchanged.}
\begin{equation}
	\begin{split}
		\bh_k = \sqrt{N} \beta_k \ba(r_k,\theta_k),
	\end{split}
\end{equation}
where $\beta_k = \frac{1}{\sqrt{4\pi r_k^2}}$ is the path loss of the LOS channel from the center of XL-FIM to user $k$.
$r_k$ and $\theta_k$ denote the distance and AOA between the center of XL-FIM and user $k$, respectively.
$\ba(r_k,\theta_k)$ denotes the steering vector between XL-FIM and user $k$ as follows
\[
\ba(r_k,\theta_k) = \frac{1}{\sqrt{N}}\left[e^{-jk_c (r_k^{(1)} - r_k)}, \dots,e^{-jk_c (r_k^{(N)} - r_k)}  \right]^{\rm T},
\]
where $k_c = 2\pi/\lambda$. $r_k^{(n)}$ denotes the distance from $n$-th radiating element of XL-FIM to user $k$.

As shown in Fig.~\ref{fig:Channel_Model}, we assume the coordinate of the $n$-th radiating element is $(x_n,y_n)$, where $x_n=\delta_n d$, $d=\frac{\lambda}{2}$, $\delta_n = n - \frac{N+1}{2},~n=1,2\dots,N$, $y_n = y_{\rm c} + \Delta y_n$, and $y_{\rm c} = 0$.
Therefore, the coordinate of the $n$-th radiating element is $( \delta_n d, \Delta y_n )$ and we have
\begin{equation}
	\begin{split}
		&r_k^{(n)} = \sqrt{ \left( r_k\cdot \sin\theta_k - x_n  \right)^2 + \left( r_k\cdot \cos\theta_k - \Delta y_n \right)^2  } \\
		&= \sqrt{ r_k^2 - 2r_k\left( x_n\sin\theta_k + \Delta y_n \cos\theta_k \right) + \left( x_n^2 + \Delta y_n^2 \right) }.
	\end{split}
\end{equation}

\vspace{-0.25cm}
Next, we define $\overline\alpha_k =  x_n\sin\theta_k + \Delta y_n \cos\theta_k $ and $\overline\beta_k = x_n^2 + \Delta y_n^2$.
By applying $\sqrt{1+x} \approx 1 + \frac{1}{2}x - \frac{1}{8} x^2 +\mathcal{O} (x^3)$, when $|x| \to$ zero, we have~\cite{Cui2022PolarCodebook}
\begin{equation}
	\begin{split}
		r_k^{(n)} = \sqrt{ r_k^2 - 2r_k\overline\alpha_k + \overline\beta_k } \approx r_k - \overline\alpha_k + \frac{\overline\beta_k - \overline\alpha_k^2}{2r_k}.
	\end{split}
\end{equation}

Then, we can obtain
\begin{equation*}
	\begin{split}
		r_k^{(n)} \approx \frac{\left( \Delta y_n \sin\theta_k - x_n \cos\theta_k \right)^2}{2r_k} -  x_n\sin\theta_k - \Delta y_n \cos\theta_k + r_k.
	\end{split}
\end{equation*}

Therefore, the steering vector corresponding to user $k$ can be reformulated as
\[
[\ba(r_k,\theta_k)]_n = \frac{ e^{-j k_c \left( \frac{\left( \Delta y_n \sin\theta_k - \delta_n d \cdot \cos\theta_k \right)^2}{2r_k} - \left( \delta_n d \cdot \sin\theta_k + \Delta y_n \cos\theta_k \right) \right) } }{\sqrt{N}}.
\]

In addition, for traditional RA, the steering vector corresponding to user $k$ is given by
\[
[\bb(r_k,\theta_k)]_n = e^{-j k_c \left( \frac{\left( \delta_n d \cdot \cos\theta_k \right)^2}{2r_k} - \delta_n d \cdot \sin\theta_k \right) } \big/ \sqrt{N}.
\]

When $r_k \rightarrow \infty$, the NF steering vector will be transformed into FF steering vector.

{\it Remark 2:}
The NF channel depends not only on the user’s angular direction but also on its position, which results in extremely high pilot overhead for channel estimation. Moreover, the use of XL-array leads to high-dimensional channel matrices, thereby imposing a substantial computational burden on channel estimation. 
To efficiently acquire the CSI of NF users, HBT schemes have been widely adopted~\cite{Zhang2025XL_RIS,Lv2024codebook,Wei2022Chinacomm}. 
However, the BT accuracy is fundamentally constrained by the beam pattern characteristics, where discrete PSs introduce strong grating lobes that significantly degrade training performance~\cite{Zhang2025NF_DPS,Zhang2025XL_RIS}. 
{\it Therefore, a critical challenge that remains to be addressed is how to suppress grating lobes and achieve accurate NF beam patterns.}

\section{NF Beam Pattern Under FIM Compensation}
Without loss of generality, we consider the beam pattern of a typical user located at $(r_u,\theta_u)$.
In order to achieve effective analysis, we first provide a definition of the beam pattern. 

\underline{\it \textbf{Definition 1}} (Beam pattern) For analog beamformer $\bv(r_u,\theta_u) \in \mathbb{C}^{N\times 1}$ corresponding to the typical user located at $(r_u,\theta_u)$, the beam pattern observed at an arbitrary location $(r,\theta)$ can be defined as follows~\cite{Cui2022PolarCodebook}
\begin{equation}\label{beam_pattern}
	\begin{split}
		&f(r_u,\theta_u,r,\theta) \triangleq \left| \bb^{\rm H}(r,\theta) \bv(r_u,\theta_u) \right| \\
		&= \frac{1}{N} \left| \sum_{n=1}^{N} e^{j\left(\angle\left([\bv(r_u,\theta_u)]_n\right) - \angle\left([\bb(r,\theta)]_n\right) \right)} \right|,~\forall r,\theta.
	\end{split}
\end{equation}
Here, replacing the $\bb(r,\theta)$ with $\ba(r,\theta)$ denotes the beam pattern under the XL-FIM scenario.

\subsection{Beam Pattern in RA-Enabled NF Scenario}
Considering traditional RA scenario, the analog beamformer with $b$-bit discrete PSs is given by
\begin{equation}
	\begin{split}
		[\bv(r_u,\theta_u)]_n = e^{j Q_b(\angle([\bb(r,\theta)]_n)},
	\end{split}
\end{equation}
where $Q_b(\angle([\bb(r,\theta)]_n)$ denotes the projection of $\angle([\bb(r,\theta)]_n)$ onto the set ${\cal C}_b$.
According to~\cite{Zhang2025XL_RIS,Zhang2024Practical_RIS}, the projection operator $Q_b(\theta)$ can be derived as
\begin{equation} \label{projection_discrete}
	Q_b(\theta)  = 
	\begin{cases}
		\frac{(z-1)\pi}{2^{b-1}}, ~\mbox{if}~ {\rm mod}(\theta,2\pi) < \frac{(2z - 1)\pi}{2^b}, \\
		\frac{z\pi}{2^{b-1}}, ~~~\, \mbox{if}~ {\rm mod}(\theta,2\pi) \geq \frac{(2z - 1)\pi}{2^b},
	\end{cases}
\end{equation}
where $z = \lfloor \frac{2^{b-1} {\rm mod}(\theta,2\pi) + \pi }{\pi} \rfloor$.

In this case, based on \eqref{beam_pattern}, \eqref{projection_discrete}, and Fourier series expansion, the beam pattern can be reformulated as~\cite{Zhang2025NF_DPS}
\begin{equation}
	\begin{split}
		&f(r_u,\theta_u,r,\theta) \triangleq \frac{1}{N} \left|\sum_{k=-\infty}^{\infty}  \Xi_k^{(b)} \sum_{n=1}^{N} e^{j\pi \delta_n \Delta_k + j \frac{\pi \lambda}{4} \delta_n^2 \Phi_k }   \right|,
	\end{split}
\end{equation}
where $\Delta_k = k\sin\theta_u - \sin\theta$ and $\Phi_k = -k\frac{\cos^2\theta_u}{r_u} + \frac{\cos^2\theta}{r}$ are defined as the {\it spatial angle difference} and the {\it ring difference}~\cite{Cui2022PolarCodebook}, respectively.
The Fourier coefficients $\Xi_k^{(b)}$ can be obtained as
\begin{equation*} 
	\Xi_k^{(b)} = 
	\begin{cases}
		\frac{2^b}{k\pi} \sin(\frac{\pi}{2^b}), ~\mbox{if}~ k \in \left\{t | t = 1 - 2^b p, p\in\mathbb{Z} \right\}, \\
		0, ~~~~~~~~~~~~\mbox{otherwise},
	\end{cases}
\end{equation*}
which can be found in \cite[Lemma 3]{Zhang2025NF_DPS}.
Therefore, it can be observed that as the PS resolution increases, the grating lobe gain gradually decreases while the main lobe gain increases. 
In this case, conventional RA are unable to effectively suppress the generation of such grating lobes, which is highly detrimental to achieving BT.

\vspace{-0.3cm}
\subsection{Beam Pattern in FIM-Enabled NF Scenario}
Our investigation of user channels in XL-FIM-enabled NF systems reveals that deformation of the FIM can introduce an additional DoF $\Delta \by = \left[\Delta y_1,\dots, \Delta y_N \right]^{\rm T}$. 
Therefore, we propose an FIM-based quantization error compensation method.

To design effective surface shape for quantization error cancellation, we first define
\begin{equation}
	\begin{split}
		\varphi_n(\Delta y_n) = &-k_c\frac{\left( \Delta y_n \sin\theta - \delta_n d  \cos\theta \right)^2}{2r} + \\
		& k_c \left( \delta_n d  \sin\theta + \Delta y_n \cos\theta \right).
	\end{split}
\end{equation}

Then, we propose the quantization error compensation condition as follows
\begin{equation} \label{compensation_condition}
	\begin{split}
		\min_{ \Delta\by, \bv(r_u,\theta_u) }&~\left\|\bv(r_u,\theta_u) - \bm\iota \right\|_2^2, \\
		{\rm s.t.}~
		&~[\bm\iota]_n = e^{j \varphi_n(\Delta y_n)},~[ \bv(r_u,\theta_u) ]_n = e^{j \vartheta_n}, \\
		&~\vartheta_n \in {\cal C}_b,\, \Delta y_n \in [-\hat y/2, \hat y/2],n=1,\dots N,
	\end{split}
\end{equation} 
where $\hat y$ is a given surface reversible deformation range.

\begin{figure}[t]	
	\centering \includegraphics[width=0.85\linewidth]{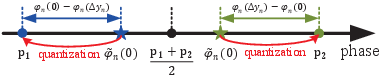}
	\vspace{-0.4cm}
	\caption{Illustration of the FIM-based quantization error compensation.}
	\vspace{-0.3cm}
	\label{fig:Quantization_compensation}
\end{figure}
\begin{figure}[t]
	\centering
	\vspace{-0.25cm}
	\subfloat[continous PS]{\includegraphics[width=0.325\columnwidth]{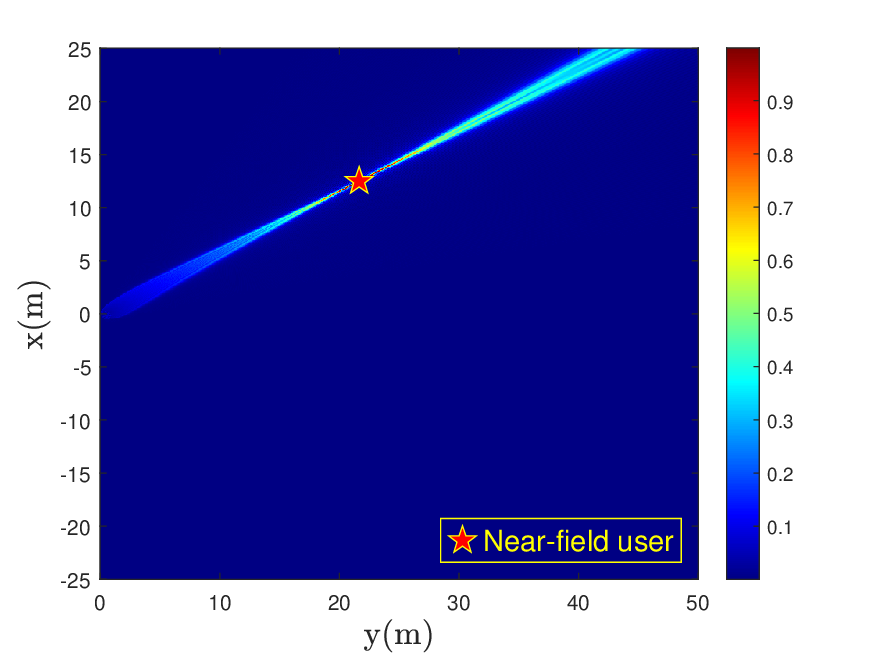}%
		\label{BeamPattern_CPS}}
	\hfil
	\subfloat[1-bit PS]{\includegraphics[width=0.325\columnwidth]{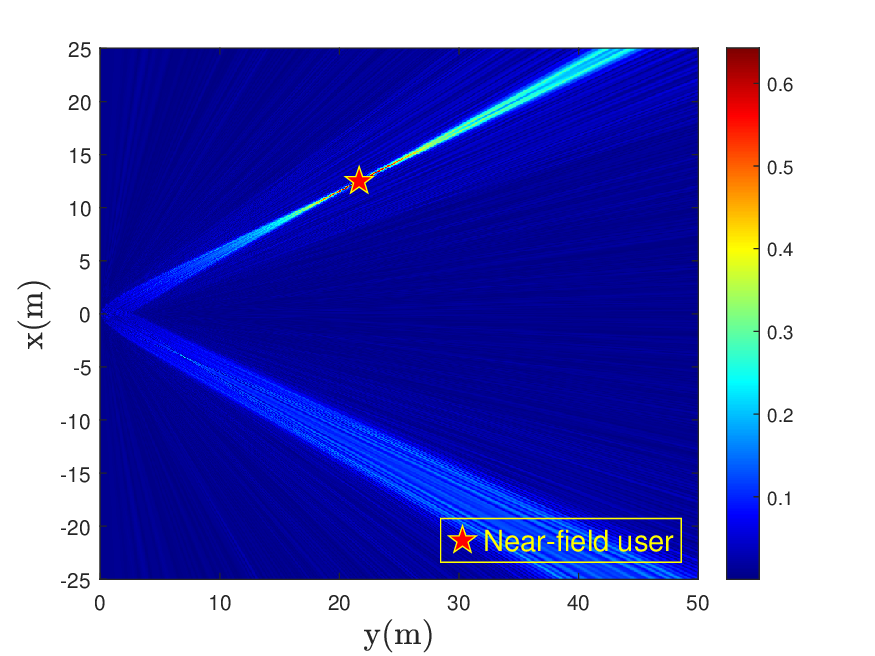}%
		\label{BeamPattern_DPS}}
	\hfil
	\vspace{-0.1cm}
	\subfloat[1-bit PS with FIM]{\includegraphics[width=0.325\columnwidth]{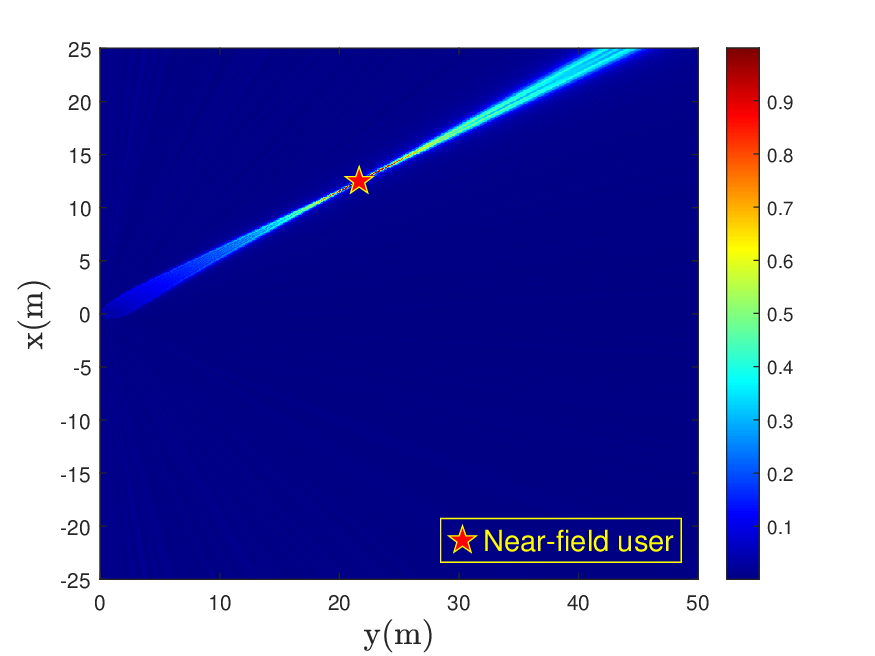}%
		\label{BeamPattern_DPS_FIM}}
	\caption{The near-field beam pattern under different phase shifters.}
	\vspace{-0.5cm}
	\label{fig:BeamPattern}
\end{figure}

To address the problem~\eqref{compensation_condition} and achieve effective elimination of quantization errors, we propose a proximal distance method to obtain the optimal surface shape.
As shown in Fig.~\ref{fig:Quantization_compensation}, points ${\rm p}_1$ and ${\rm p}_2$ denote the phase points at discrete phase set ${\cal C}_b $.
Define $\widetilde\varphi_n(0) = {\rm mod}\left(\varphi_n(0), 2\pi\right)$, and $Q_b\left( \widetilde\varphi_n(0) \right) $ denotes the projection of $\widetilde\varphi_n(0)$ onto the set ${\cal C}_b$.
When ${\rm p}_1 \leq \widetilde\varphi_n(0) \leq \frac{{\rm p}_1 + {\rm p}_2}{2} $, $Q_b\left( \widetilde\varphi_n(0) \right) = {\rm p}_1$ and otherwise $Q_b\left( \widetilde\varphi_n(0) \right) = {\rm p}_2$.
Therefore, we can achieve quantization error compensation by the following equation
\begin{equation} \label{error_compensation}
	\begin{split}
		\left| \varphi_n(\Delta y_n) - \varphi_n(0) \right| = \left| \widetilde\varphi_n(0) - Q_b\left( \widetilde\varphi_n(0) \right) \right|.
	\end{split}
\end{equation}

To ensure that equation~\eqref{error_compensation} holds, we next discuss the optimal surface shape design under different cases.

\subsubsection{$\theta \neq 0$}
In this case, we have
\begin{equation}
	\begin{split}
		\varphi_n(\Delta y_n) - \varphi_n(0) = a_{1,n} \Delta y_n^2 + a_{2,n}\Delta y_n,
	\end{split}
\end{equation}
where $a_{1,n} = \frac{-k_c \sin^2\theta}{2 r}$ and $a_{2,n} = \frac{k_c\cos\theta( r + \delta_n d \sin\theta) }{ r }$.

Define $a_{3,n} = \left| \widetilde\varphi_n(0) - Q_b\left( \widetilde\varphi_n(0) \right) \right|$, we examine the following different cases.

\noindent $\blacklozenge$ {\bf{Case 1:}} When $\widetilde\varphi_n(0) \leq Q_b\left( \widetilde\varphi_n(0) \right)$, we have
\begin{equation}
	\begin{split}
		a_{1,n} \Delta y_n^2 + a_{2,n}\Delta y_n - a_{3,n} = 0,~n=1,2,\dots,N.
	\end{split}
\end{equation}

Therefore, the optimal surface shape can be derived as
\begin{equation}
	\begin{split}
		\Delta y_n = \frac{-a_{2,n} + \sqrt{a_{2,n}^2 + 4a_{1,n}a_{3,n}} }{2a_{1,n}},~n=1,2,\dots,N.
	\end{split}
\end{equation}

\noindent $\blacklozenge$ {\bf{Case 2:}} When $\widetilde\varphi_n(0) > Q_b\left( \widetilde\varphi_n(0) \right)$, we have
\begin{equation}
	\begin{split}
		a_{1,n} \Delta y_n^2 + a_{2,n}\Delta y_n + a_{3,n} = 0,~n=1,2,\dots,N.
	\end{split}
\end{equation}

Therefore, the optimal surface shape can be derived as
\begin{equation}
	\begin{split}
		\Delta y_n = \frac{-a_{2,n} + \sqrt{a_{2,n}^2 - 4a_{1,n}a_{3,n}} }{2a_{1,n}},~n=1,2,\dots,N.
	\end{split}
\end{equation}

\subsubsection{$\theta = 0$}
In this case, we have
\begin{equation}
	\begin{split}
		\varphi_n(\Delta y_n) - \varphi_n(0) = k_c\Delta y_n.
	\end{split}
\end{equation}

Then, we also have the following two cases.

\noindent $\blacklozenge$ {\bf{Case 1:}} When $\widetilde\varphi_n(0) \leq Q_b\left( \widetilde\varphi_n(0) \right)$, we have
\begin{equation}
	\begin{split}
		k_c \Delta y_n - a_{3,n} = 0,~n=1,2,\dots,N.
	\end{split}
\end{equation}

Therefore, the optimal surface shape can be derived as
\begin{equation}
	\begin{split}
		\Delta y_n = \frac{a_{3,n}}{k_c},~n=1,2,\dots,N.
	\end{split}
\end{equation}

\noindent $\blacklozenge$ {\bf{Case 2:}} When $\widetilde\varphi_n(0) > Q_b\left( \widetilde\varphi_n(0) \right)$, we have
\begin{equation}
	\begin{split}
		k_c \Delta y_n + a_{3,n} = 0,~n=1,2,\dots,N.
	\end{split}
\end{equation}

Therefore, the optimal surface shape can be derived as
\begin{equation}
	\begin{split}
		\Delta y_n = -\frac{a_{3,n}}{k_c},~n=1,2,\dots,N.
	\end{split}
\end{equation}

To more clearly demonstrate the effectiveness of the FIM in mitigating quantization errors, we consider an example where $\theta_u$ takes values of $\pi/6$, the propagation distance $r$ is 25 meters, and the array size $N$ is 512. 
The resulting beam patterns for continuous PS, 1-bit discrete PS, and 1-bit discrete PS with FIM-based compensation are depicted in Fig.~\ref{fig:BeamPattern}.
As can be observed, surface shaping enabled by the FIM can effectively suppress grating lobes, achieving beam patterns comparable to those obtained with continuous PS.

\begin{figure}[t]
	\centering
	\vspace{-0.25cm}
	\subfloat[$B_{\rm y}=2$]{\includegraphics[width=0.5\columnwidth]{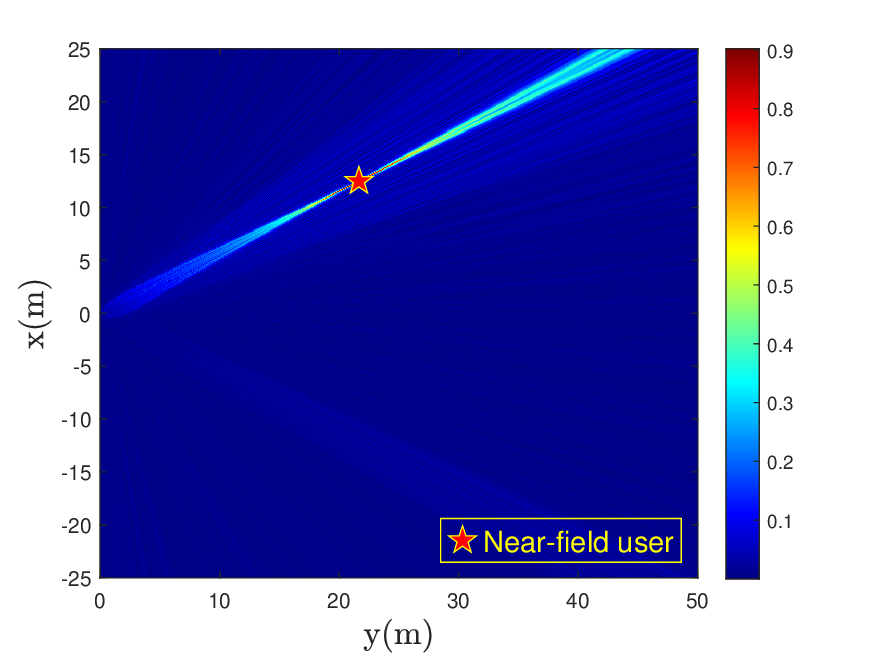}%
		\label{Discrete_FIM_beampattern_2bit}}
	\hfil
	\subfloat[$B_{\rm y}=3$]{\includegraphics[width=0.5\columnwidth]{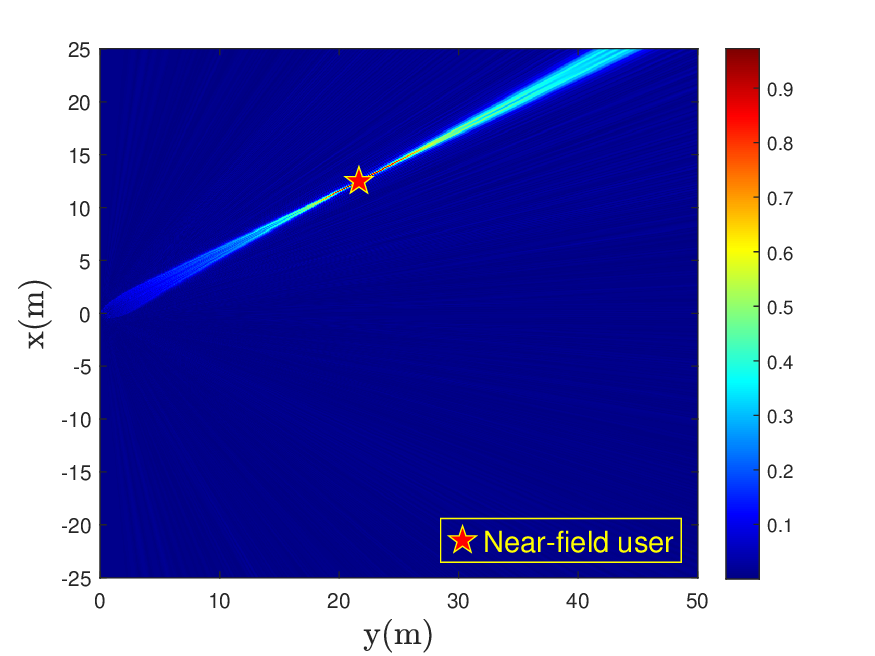}%
		\label{Discrete_FIM_beampattern_3bit}}
	\caption{The near-field beam pattern under the discrete position condition of the FIM element, $r_u = 25\,$m, $N=512$.}
	\vspace{-0.2cm}
	\label{fig:Discrete_FIM_beampattern}
\end{figure}

{\it Remark 3:} 
The above solution yields the continuously optimized surface shape that can effectively compensate for the phase-quantization errors, provided that the required element displacements lie within the feasible shaping range. 
In practice, the shaping range of an FIM is typically on the order of the wavelength~\cite{Ni2022NC,Bai2022FIM,An2025FIM,An2025FIM_TCOM}. 
As will be shown through the following dominant-order analysis, such an inherent shaping capability is generally sufficient to realize the required displacement adjustment.
Moreover, practical surface shaping mechanisms may restrict each element to a finite number of discrete positions.
Let $B_{\mathrm y}$ denote the FIM shaping-position quantization resolution.
For $2^{B_{\mathrm y}}$ uniformly spaced positions including both endpoints, the maximum position-quantization error satisfies $|e_n| \leq \hat y / \left(2^{B_{\mathrm y}+1}-2\right)$.
According to the channel phase model, the resulting residual phase error is bounded by
\[
	\left|\varphi_n(\Delta y_n^\star + e_n) - \varphi_n(\Delta y_n^\star)\right|
	\leq
	\frac{k_c\hat y \zeta_n}{2\left(2^{B_{\mathrm y}}-1\right)} 
	+
	\frac{k_c\sin^2\theta\hat y^2}
	{8r\left(2^{B_{\mathrm y}}-1\right)^2}.
\]
where $\zeta_n = \left|
\cos\theta-
\frac{\sin\theta\left(\Delta y_n^\star\sin\theta-x_n\cos\theta\right)}{r}
\right|$ and $\Delta y_n^\star$ is the continuously optimized displacement.
Hence, continuous morphing provides an ideal compensation benchmark, whereas discrete shaping results in bounded residual phase errors. The above bound decreases as the number of feasible positions increases, indicating that a finer morphing-position resolution enables more accurate compensation and gradually approaches the continuous-shaping performance, as shown in Fig.~\ref{fig:Discrete_FIM_beampattern}.

\subsection{Error-Compensation-Based Shaping Range Analysis}
In this section, we aim to determine the order of magnitude of the FIM shaping range required to effectively compensate for phase-shift quantization errors.
For the $n$-th FIM radiating element, shifting its position by $\Delta y_n$ induces the following phase difference
\begin{equation}
	\begin{split}
		\varphi_n(\Delta y_n) - \varphi(0) = k_c \left( \omega_n \Delta y_n - \sin^2\theta\Delta y_n^2 / 2r \right),
	\end{split}
\end{equation}
where $\omega_n = \cos\theta\left( 1 + \frac{\delta_n d \sin\theta}{r} \right) $.

For $b$-bit PS, the corresponding maximum equivalent path-length error is $\lambda/2^{b+1}$. 
To account for all possible phase quantization errors, it suffices to consider the worst-case equation, given by $\omega_n \Delta y_n - \sin^2\theta\Delta y_n^2 / 2r = \lambda/2^{b+1}$.
Therefore, we obtain the following result
\begin{equation}
	\begin{split}
		\left| \Delta y_{n,{\rm req}} \right| = \left| u_{n,{\rm req}}(r,\theta) \right| \cdot \lambda,
	\end{split}
\end{equation}
where $u_{n,{\rm req}}(r,\theta) = 1/\left[2^b \left(\omega_n + \sqrt{\omega_n^2 - \lambda \sin^2\theta / (2^b r) } \right) \right]$.

Then, the shaping range can be scaled as follows
\begin{equation*} 
	\begin{cases}
		\left| \Delta y_{n,{\rm req}} \right| \leq \frac{\lambda}{2^b \omega_n} \leq \frac{\lambda}{2^b  \cos\theta }  , ~~~~~~~~~~~~~~~~\,\,\,\, \mbox{if}~ \theta \geq 0, \\
		\left| \Delta y_{n,{\rm req}}  \right| \leq \frac{\lambda}{2^b \omega_n} \leq  \frac{4 r_{\rm min} \lambda}{2^b \cos\theta (4 r_{\rm min} - (N-1)\lambda)}, ~\mbox{if}~ \theta < 0,
	\end{cases}
\end{equation*}
where $r_{\rm min}$ denotes the minimum communication distance considered, and the second inequality can be approximated as $\left| \Delta y_{n,{\rm req}}  \right| \leq \frac{\lambda}{2^{b-1}  \cos\theta }$ when $N\leq 2048$ and $\lambda=0.01\,$m.

Furthermore, in practical mmWave and THz arrays, the effective azimuth FoV is typically restricted to approximately $[-\pi/3,\pi/3]$ due to element-pattern attenuation and scan loss, thereby maintaining sufficient array gain while suppressing grating lobes. This FoV assumption has been widely adopted in both academic studies and industry specifications for near-field array modeling~\cite{Natarajan2013AWPL,Jin2021TAP}.
Accordingly, a key conclusion is that, under these conditions, an FIM shaping range on the order of the wavelength, i.e., $\mathcal{O}(\lambda)$, is sufficient to effectively compensate for phase quantization errors.

It should be emphasized that the above result is a conditional uniform bound over the considered operating region rather than an unconditional global guarantee for arbitrary user locations and array configurations. 
In particular, the required shaping range may exceed the wavelength scale for near-endfire users, extremely large apertures, or communication distances below the prescribed $r_{\rm min}$.


\begin{figure}[t]	
	\centering \includegraphics[width=0.8\linewidth]{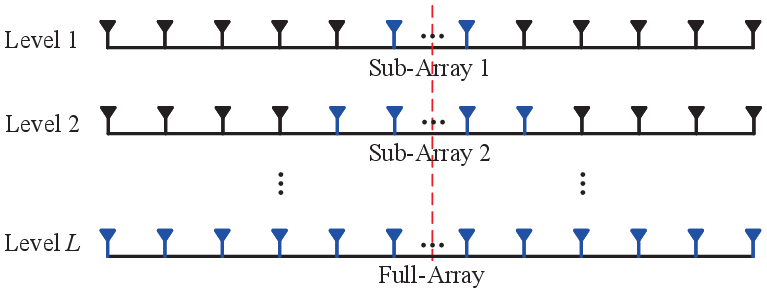}
	\vspace{-0.1cm}
	\caption{The hierarchical sub-array structure in the proposed HSABT strategy.}
	\vspace{-0.5cm}
	\label{fig:Hierarchical_Sub_Array}
\end{figure}
\section{Hierarchical Sub-Array Beam Training}
In this section, we aim to efficiently capturing main path of NF users through BT, leveraging the FIM-based quantization error compensation method. 
By compensating for the quantization errors using the FIM, the BT process becomes effectively equivalent to that of a system with continuous PS.
Let us define $\phi_u = \sin\theta_u\in[-1,1]$, whereby the NF channel can be equivalently reformulated as
\begin{equation}
	\begin{split}
		[\ba(r_u,\phi_u)]_n = e^{-j k_c \left( \frac{ \delta_n^2 d^2 (1-\phi_u^2) }{2r_u} - \delta_n d \phi_u \right) } \big/ \sqrt{N}.
	\end{split}
\end{equation}

\subsection{Sub-Array Beam Sweeping}
In this section, we develop a BT strategy based on the equivalent channel model and the continuous PS. 
Departing from conventional approaches such as polar codebook and DFT codebook, we develop an efficient BT method with low overhead, referred to as {\it Hierarchical Sub-Array Beam Training (HSABT)}. 
As shown in Fig.~\ref{fig:Hierarchical_Sub_Array}, the HSABT method begins by partitioning the array into small central sub-arrays, each performing coarse BT to identify potential directions. 
At successive levels, the sub-array size is progressively increased, thereby refining the beam pattern resolution. 
This hierarchical structure not only substantially lowers the overall pilot overhead compared with exhaustive DFT or polar codebook searches but also exploits the sub-array architecture to improve training accuracy in both angular and radial dimensions.
{\it However, this approach also raises two critical challenges, i.e., determining how to partition the sub-arrays at each hierarchical level and how to efficiently perform the corresponding beam sweeping}.

The key to overcoming these two challenges lies in exploring the beam space characteristics corresponding to different numbers of radiating elements. 
Accordingly, we proceed to analyze the beam space properties of the equivalent system, focusing on the beam-width and beam-depth, whose definitions are provided below.

\underline{\it \textbf{Definition 2}} (Beam-width) The beam-width quantifies the angular span of the beam pattern within the {\it distance-ring} (which is defined as $\frac{1 - \phi^2}{r}$~\cite{Cui2022PolarCodebook}), over which the beam power decreases from its peak value to zeros.
Therefore, the beam-width refers to the null-to-null angular range of the beam power (from $\phi_{\rm right}$ to $\phi_{\rm left}$), which is given by
\begin{equation}
	\begin{split}
		{\rm BW} \triangleq \left| \phi_{\rm right} - \phi_{\rm left} \right|.
	\end{split}
\end{equation}

\underline{\it \textbf{Definition 3}} (Beam-depth) The beam-depth characterizes the distance range along a given observation angle $\phi$ over which the beam power drops from its peak value to half, i.e., $r\in [r_{\rm small}, r_{\rm large}]$. Consequently, the beam-depth can be mathematically expressed as
\begin{equation}
	\begin{split}
		{\rm BD} \triangleq \left| r_{\rm large} - r_{\rm small} \right|.
	\end{split}
\end{equation}

\subsubsection{Sub-Array Partitioning Based on Beam-Width}
Based on {\it Definition 1}, in the {\it distance-ring}, the beam pattern can be equivalently calculated as
\begin{equation}\label{beam_pattern_sinc}
	\begin{split}
		&f(r_u,\phi_u,r,\phi) = \left| \frac{\sin\left(\frac{N\pi}{2}(\phi_u - \phi) \right)}{N\sin\left(\frac{\pi}{2}(\phi_u - \phi) \right)} \right|.
	\end{split}
\end{equation}

\begin{figure}[t]
	\vspace{-0.2cm}	
	\centering \includegraphics[width=0.8\linewidth]{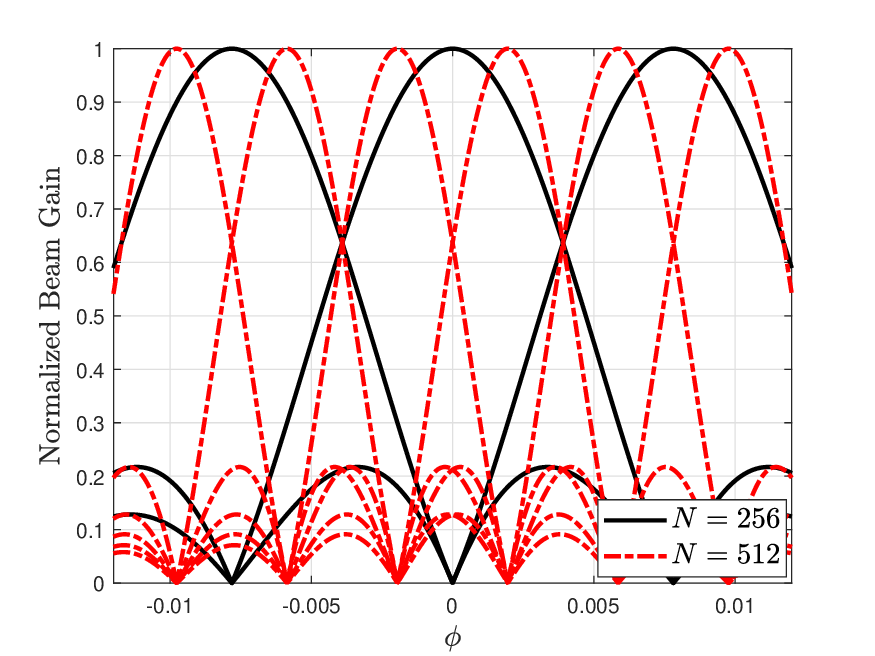}
	\vspace{-0.3cm}
	\caption{The beam pattern obtained in different number of radiating elements.}
	\vspace{-0.5cm}
	\label{fig:Angle_sample}
\end{figure}

Based on this, we can obtain the beam-width as $4/N$.
This conclusion is particularly encouraging, as it reveals that doubling the number of radiating elements in the array exactly halves the beam-width. 
As a result, when the number of radiating elements at level $l+1$ is twice that of level $l$, the coverage of two beams at level $l+1$ perfectly coincides with the coverage of a single beam at level $l$.
Similarly, this conclusion also holds for FF channels.
To validate this finding, we simulate the beam patterns for $N=256$ and $N=512$, respectively. 
As shown in Fig.~\ref{fig:Angle_sample}, the single beam for $N=256$ precisely covers the two beams for $N=512$.

Based on the above analysis, we propose an efficient sub-array partitioning method. 
Specifically, we set the number of radiating elements in the level-1 sub-array to $M$, so that the number of elements in the level-$l$ sub-array becomes $M\times 2^{l-1}$. 
This hierarchical sub-array approach enables accurate BT in the angular domain.
The angular-domain BT overhead for the first level is $N_1$. 
{\it For each subsequent sub-array level, BT in the angular domain only requires two additional scans based on the previous level. }
As the number of radiating elements in each sub-array increases, NF effects become increasingly significant. 
For instance, at a carrier frequency of 30 GHz with $N=512$, where the first-level sub-array consists of 16 radiating elements, the BT at the fifth and sixth levels must consider the coupling between the distance and angular domains. 
This leads to a key technical challenge: {\it how to incorporate distance-domain stratification into the hierarchical sub-array-based angular-domain BT.}


\vspace{0.1cm}
\subsubsection{Distance Partitioning Based on Beam-Depth}
To address the above challenge, we investigate the concept of beam-depth and the corresponding sampling criterion in the distance domain, aiming to achieve full-coverage hierarchical distance design.
Due to the extremely large number of radiating elements, for a given $\phi_u$, the beam pattern can be approximated by that of a continuous-aperture array based on {\it Definition 3}, i.e.,
\begin{equation}\label{NF_beam_gain}
	\begin{split}
		f(r_u,\phi_u,r,\phi) \approx \left| \frac{1}{Nd} \int_{-\frac{Nd}{2}}^{\frac{Nd}{2}} e^{j \Delta_r n^2} \,dn \right| = \left| \frac{C(\kappa) + jS(\kappa) }{\kappa} \right|,
	\end{split}
\end{equation}
where $\Delta_r = \frac{k_c(1-\phi_u^2)(r-r_u)}{2r_u r}$ and $\kappa = \sqrt{\frac{N^2 d^2 (1 - \phi_u^2)}{2\lambda}\left| \frac{1}{r_u} - \frac{1}{r} \right| }$.
$C(\kappa) = \int_{0}^{\kappa} \cos(\frac{\pi}{2}t^2)\,dt$ and $S(\kappa) = \int_{0}^{\kappa} \sin(\frac{\pi}{2}t^2)\,dt$ are Fresnel functions~\cite{Cui2022PolarCodebook,Bjornson2021NF}.

Therefore, the condition under which the beam gain decreases to half of its peak value can be expressed as follows
\begin{equation}
	\begin{split}
		\left| \frac{C(\kappa) + jS(\kappa) }{\kappa} \right| = 0.5,
	\end{split}
\end{equation}
which is equivalent to
\begin{equation} \label{ring_difference}
	\begin{split}
		\left| \frac{1}{r_u} - \frac{1}{r} \right| = \frac{2\lambda\kappa^2}{N^2 d^2 (1-\phi_u^2)} \triangleq \frac{1}{Z_{\Delta}(\phi_u)},
	\end{split}
\end{equation}
where $\kappa \approx 1.56$ can be observed from the beam pattern.

Based on~\eqref{ring_difference}, we can derive the beam-depth as
\begin{equation} \label{beam_depth}
	{\rm BD}(r_u)  = 
	\begin{cases}
		\frac{2r_u^2 Z_{\Delta}(\phi_u) }{Z_{\Delta}^2(\phi_u) - r_u^2}, ~\mbox{if}~r_u < Z_{\Delta}(\phi_u) , \\
		\infty, ~~~~~~~~~~\,\mbox{if}~r_u \geq Z_{\Delta}(\phi_u).
	\end{cases}
\end{equation}

Although the beam-depth can be determined, unfortunately, there is no explicit multiple-coverage relationship between the beam-depths of different sub-array levels, which makes it difficult to guide HBT in the distance domain. 
However, from another perspective, the beam-depth provides inspiration for the sampling strategy in the distance domain, denoted as 
\begin{equation}
	\begin{split}
		r_z = \frac{1}{z}Z_{\Delta}(\phi),~z = 1,2,\dots,Z,
	\end{split}
\end{equation}
which is fully consistent with the sampling method in~\cite{Cui2022PolarCodebook}. 
This naturally raises the question of whether a proximal distance observation-based approach, similar to that used for determining $\kappa$, can be employed to guide the hierarchical design in the distance domain.

\begin{figure}[t]
	\vspace{-0.3cm}	
	\centering \includegraphics[width=0.8\linewidth]{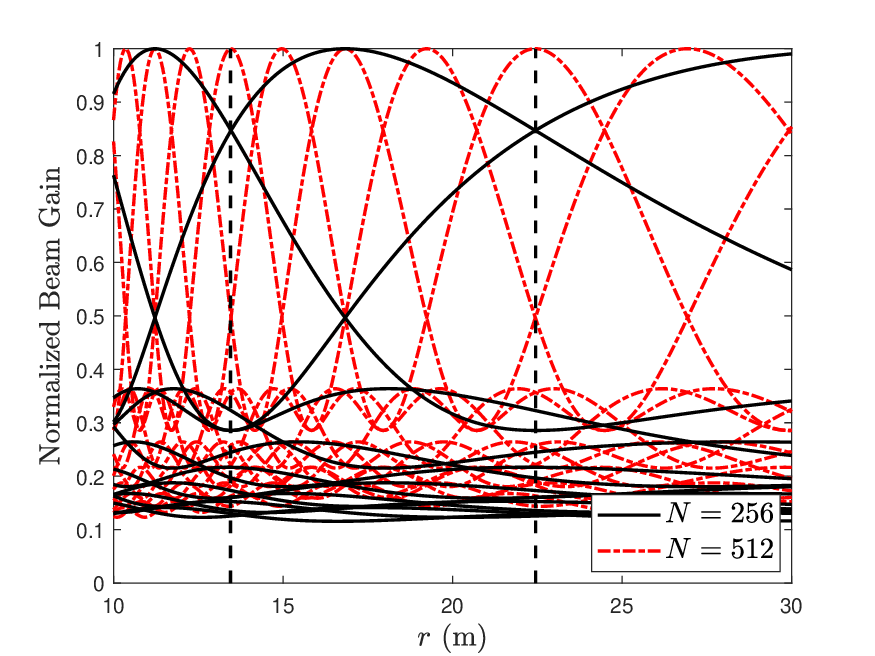}
	\vspace{-0.4cm}
	\caption{The beam pattern obtained in different number of radiating elements and range samples, $\phi = 0$.}
	\vspace{-0.5cm}
	\label{fig:Range_sample}
\end{figure}

Motivated by this, in Fig.~\ref{fig:Range_sample}, we present the beam patterns corresponding to the sampling points within the $[10, 30]\,$m range for $N=256$ and $N=512$, respectively. 
Encouragingly, we observe that for $N=256$, each BT instance corresponds to a single distance sampling point. 
By calculating the coverage range defined by the intersections of the beam patterns at this sampling point with those at its two neighboring points, we can effectively determine the sampling points that need BT for $N=512$ (the next level sub-array). 
This approach thus enables efficient HBT in the distance domain.

Therefore, in the following, we will derive the distance range between the two intersection points of the beam patterns corresponding to three adjacent distance sampling points.
We consider $2\leq z \leq Z-1$, the steering vectors corresponding to sample points are given by
\begin{equation}
	\begin{split}
		& \left[ \ba(r_{z-1},\phi) \right] = e^{-j k_c \left( \frac{ \delta_n^2 d^2 (1-\phi^2) }{2r_{z-1}} - \delta_n d \phi \right) } / \sqrt{N}, \\
		& \left[ \ba(r_{z},\phi) \right] = e^{-j k_c \left( \frac{ \delta_n^2 d^2 (1-\phi^2) }{2r_{z}} - \delta_n d \phi \right) } / \sqrt{N}, \\
		& \left[ \ba(r_{z+1},\phi) \right] = e^{-j k_c \left( \frac{ \delta_n^2 d^2 (1-\phi^2) }{2r_{z+1}} - \delta_n d \phi \right) } / \sqrt{N}.
	\end{split}
\end{equation}

We then aim to calculate a variable $r_1$ subject to
\begin{equation}
	\begin{split}
		\left| \ba^{\rm H}(r_1,\phi) \ba(r_{z-1},\phi) \right| = \left| \ba^{\rm H}(r_1,\phi) \ba(r_{z},\phi) \right|.
	\end{split}
\end{equation}

Similarly, we also aim to calculate a variable $r_2$ subject to
\begin{equation}
	\begin{split}
		\left| \ba^{\rm H}(r_2,\phi) \ba(r_{z},\phi) \right| = \left| \ba^{\rm H}(r_2,\phi) \ba(r_{z+1},\phi) \right|.
	\end{split}
\end{equation}

As a result, for $r_z$, we have
\begin{equation}
	\begin{split}
		&\ba^{\rm H}(r_1,\phi) \ba(r_{z},\phi) \\
		&= \frac{1}{N} \sum_{n=1}^{N} e^{-j k_c \frac{ \delta_n^2 d^2 (1-\phi^2) }{2r_{1}} + j k_c \frac{ \delta_n^2 d^2 (1-\phi^2) }{2r_{z}} } \\
		&= \frac{1}{N} \sum_{n=1}^{N} e^{-j \varrho \delta_n^2 \Delta_z(r_1)},
	\end{split}
\end{equation}
where $\Delta_z(r_1) = \frac{1}{r_1} - \frac{1}{r_z}$ and $\varrho = \frac{k_c d^2 (1-\phi^2)}{2}$.

Subsequently, we can calculate the beam gain as
\begin{equation}
	\begin{split}
		&\left| \ba^{\rm H}(r_1,\phi) \ba(r_{z},\phi) \right| = \left| \frac{1}{N} \sum_{n=1}^{N} e^{-j \varrho \delta_n^2 \Delta_{z}(r_1)} \right|,  \\
		&\left| \ba^{\rm H}(r_1,\phi) \ba(r_{z-1},\phi) \right| = \left| \frac{1}{N} \sum_{n=1}^{N} e^{-j \varrho \delta_n^2 \Delta_{z-1}(r_1)} \right|.
	\end{split}
\end{equation}

We aim to achieve the following equation
\begin{equation} \label{intersection_point}
	\begin{split}
		\left| \frac{1}{N} \sum_{n=1}^{N} e^{-j \varrho \delta_n^2 \Delta_{z}(r_1)} \right| = \left| \frac{1}{N} \sum_{n=1}^{N} e^{-j \varrho \delta_n^2 \Delta_{z-1}(r_1)} \right|.
	\end{split}
\end{equation}

Since the reference point is the center point of the FIM, we can express~\eqref{intersection_point} equivalently as follows
\begin{equation}
	\begin{split}
		\left| \Delta_{z}(r_1) \right| = \left| \Delta_{z-1}(r_1) \right|.
	\end{split}
\end{equation}

Due to $\frac{1}{r_1} - \frac{1}{r_{z-1}} > 0$ and $\frac{1}{r_1} - \frac{1}{r_{z}} < 0$, we have
\begin{equation}
	\begin{split}
		r_1 = \frac{2r_{z-1} r_z}{r_{z-1} + r_z} = \frac{Z_\Delta(\phi)}{z - 1/2}.
	\end{split}
\end{equation}

Similarly, we can derive $r_2$ as
\begin{equation}
	\begin{split}
		r_2 = \frac{Z_\Delta(\phi)}{z + 1/2}.
	\end{split}
\end{equation}

Therefore, the sampling points for the next level should be selected within the following range
\[
\left[\frac{Z_\Delta(\phi)}{z + 1/2}, \frac{Z_\Delta(\phi)}{z - 1/2} \right],~z = 1,2,\dots,Z.
\]

\subsubsection{Beam Pattern Intersection-Based Beam Training}
In general, as the number of arrays radiating elements increases, the angular and distance resolutions achieved by BT become higher. For example, over the entire angular domain $[-1,1]$, existing methods, such as the DFT codebook and polar codebook, can distinguish $N$ angular directions. However, we aim to break through this conventional resolution limit for a given $N$. To this end, we propose a beam pattern intersection-based beam training method, with the objective of improving the resolutions in both the angular and distance domains. For clarity, we use $N=256$ as an illustrative example.

\begin{figure}[t]
	\vspace{-0.3cm}	
	\centering \includegraphics[width=0.8\linewidth]{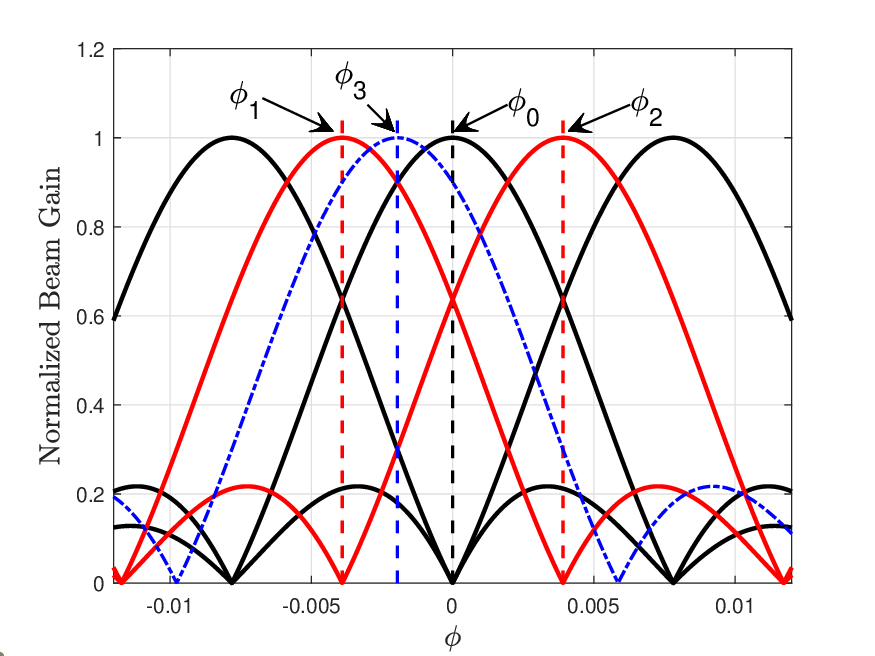}
	\vspace{-0.4cm}
	\caption{An example of the beam training method based on beam pattern intersections in the angular domain for $N=256$.}
	\vspace{-0.2cm}
	\label{fig:Angle_Intersection}
\end{figure}

As shown in Fig.~\ref{fig:Angle_Intersection}, hierarchical sub-array beam scanning in the angular domain first yields a candidate user angle of $\phi_0$, indicating that the user lies within the coverage region of the beam whose peak direction is $\phi_0$. To further improve the angular resolution, we use the intersection points between the beam corresponding to $\phi_0$ and its two adjacent beams as sampling points to generate two new beams, denoted by $\phi_1$ and $\phi_2$. Subsequently, two beam scans are performed using the beams corresponding to $\phi_1$ and $\phi_2$. If the beam associated with $\phi_1$ achieves a higher scanning gain, the user angle can be determined to lie between $\phi_0$ and $\phi_1$. In this case, a new candidate user angle can be obtained as $\phi_3=(\phi_0+\phi_1)/2$, thereby doubling the angular resolution.
\begin{figure}[t]
	\vspace{-0.3cm}	
	\centering \includegraphics[width=0.8\linewidth]{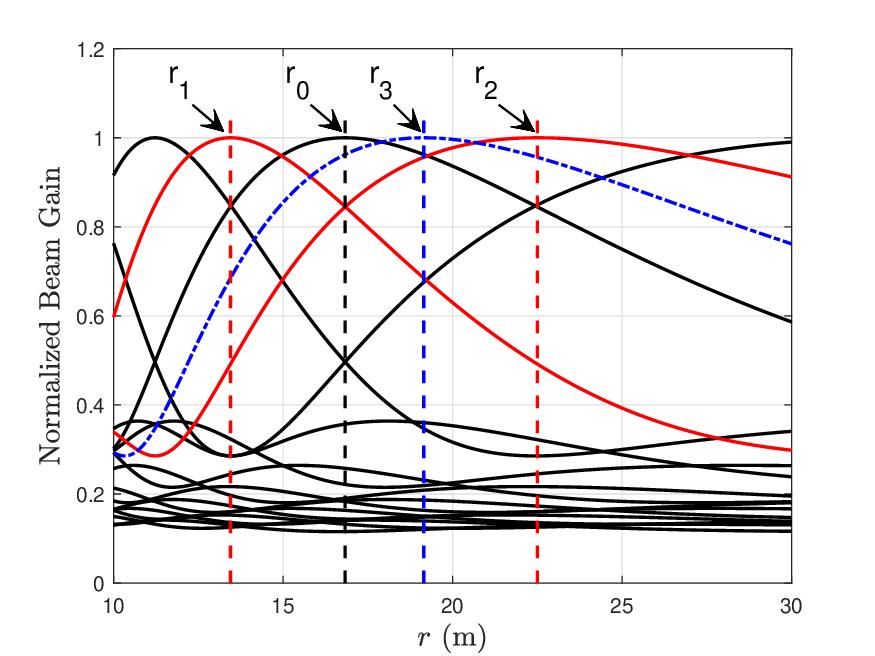}
	\vspace{-0.4cm}
	\caption{An example of the beam training method based on beam pattern intersections in the distance domain for $N=256$.}
	\vspace{-0.5cm}
	\label{fig:Distance_Intersection}
\end{figure}

As shown in Fig.~\ref{fig:Distance_Intersection}, hierarchical sub-array beam scanning in the distance domain first yields a candidate user distance of $r_0$, indicating that the user lies within the coverage region of the beam whose peak distance is $r_0$. To further improve the distance resolution, we use the intersection points between the beam corresponding to $r_0$ and its two adjacent beams as sampling points to generate two new beams, denoted by $r_1$ and $r_2$. Subsequently, two beam scans are performed using the beams corresponding to $r_1$ and $r_2$. If the beam associated with $r_2$ achieves a higher scanning gain, the user distance can be determined to lie between $r_0$ and $r_2$. In this case, a new candidate user distance can be obtained as $r_3$, thereby improving the distance resolution.

\subsubsection{NF and FF Beam Switching}
The conventional Rayleigh distance provides a geometric boundary for distinguishing the NF and FF regions~\cite{Tarboush2024Cross_Field}. 
However, such a predefined propagation boundary does not necessarily indicate which type of codeword can achieve the highest beamforming gain for a specific user. 
In particular, users located around the near-far transition region may still be effectively served by either an NF focusing codeword or an FF angular codeword. 
Therefore, instead of directly determining the propagation region according to the Rayleigh distance, we propose a measurement-driven NF and FF beam switching mechanism, in which a small number of candidate NF codewords and an FF codeword are jointly evaluated according to their measured beamforming gains.

\begin{figure}[t]
	\vspace{-0.3cm}	
	\centering \includegraphics[width=0.8\linewidth]{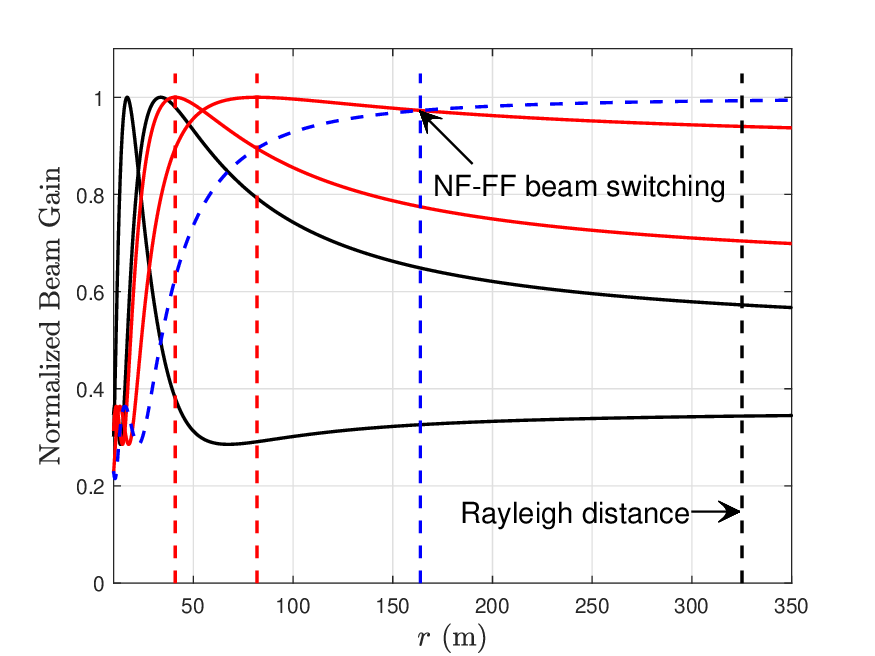}
	\vspace{-0.4cm}
	\caption{An example of the beam training method based on beam pattern intersections in the distance domain for $N=256$.}
	\vspace{-0.5cm}
	\label{fig:Hybrid_NF_BT}
\end{figure}

For a user located in the angular direction $\phi$, the normalized beamforming gain achieved by the corresponding FF codeword can be characterized by~\eqref{NF_beam_gain}.
Specifically, the parameter $\kappa$ is given by
\begin{equation}
	\begin{split}
		\kappa = \sqrt{\frac{N^2d^2(1-\phi^2)}{2\lambda} \left| \frac{1}{r_c} - \frac{1}{r} \right| },
	\end{split}
\end{equation}
where $r_c$ denotes the focal distance of the adopted codeword and $r$ denotes the actual user distance.
For an FF codeword, the focal distance tends to infinity, i.e., $r_c \rightarrow \infty$.
Therefore, $\kappa$ reduces to $\kappa_{\rm FF} = \sqrt{N^2d^2(1-\phi^2)/(2\lambda r)}$.
As shown in Fig.~\ref{fig:Hybrid_NF_BT}, the blue dashed curve represents the beam gain obtained by users at different distances when the transmitter uses FF codewords.
It can be observed that matching between the FF codeword and the actual channel improves as the distance $r$ increases.
Consequently, the corresponding beamforming gain approaches the maximum value in the sufficiently large-distance regime.
To determine the range in which both NF and FF codewords should be considered, we introduce a normalized gain threshold $\xi \in (0,1)$.
Let $\kappa_\xi$ satisfy $\left| \frac{C(\kappa_\xi) + jS(\kappa_\xi) }{\kappa_\xi} \right| = \xi$.
The corresponding user distance is therefore given by $r_{\xi}(\phi) = N^2d^2(1-\phi^2)/2\lambda \kappa_\xi^2$.
When $r \geq r_{\xi}(\phi)$, the FF codeword can achieve at least a $\xi$-fraction of the maximum normalized beamforming gain.
It should be emphasized that $r_{\xi}(\phi)$ is used only to determine the candidate transition region rather than serving as the final NF and FF switching boundary.
In this paper, we set $\xi = 0.9$.
According to the normalized gain curve, a 90\% normalized gain approximately corresponds to $\kappa_\xi=1$.
Hence, the corresponding distance becomes $r_{0.9}(\phi) = N^2d^2(1-\phi^2)/2\lambda$.

Recall that the beam-depth-based distance threshold is $Z_{\Delta}(\phi) = N^2d^2(1-\phi^2)/(2\lambda (1.56)^2)$.
Therefore, we can obtain $r_{0.9}(\phi) = 1.56^2 Z_{\Delta}(\phi)$.
To further characterize this transition region, additional NF candidate codewords are constructed at
\begin{equation}
	\begin{split}
		\tilde r_{\tilde z}(\phi) = \frac{N^2d^2(1-\phi^2)}{2\lambda \tilde z},~~\tilde z \in \mathbb{Z}^+.
	\end{split}
\end{equation}

To ensure that the additional sampling points remain beyond the original beam-depth threshold, we require $\tilde r_{\tilde z}(\phi) > Z_{\Delta}(\phi)$.
Therefore, we can derive $\tilde z \leq 2$.
Therefore, at most two additional NF candidate codewords are required to characterize the near-far transition region. This significantly limits the additional BT overhead.
Finally, the candidate NF codewords and the FF codeword aligned with the estimated angular direction are jointly evaluated through BT.
Let $ {\cal Q}_{\rm NF} = \left\{ \bq_{\rm NF}(\phi, \tilde r_{\tilde z}): \tilde z \leq 2 \right\} $ denote the valid set of additional NF candidate codewords, and let $\bq_{\rm FF}(\phi)$ denote the corresponding FF codeword. 

Subsequently, the final NF or FF beam selection is determined by the actual measured beamforming gain rather than by a predefined Rayleigh distance threshold.
Specifically, the NF codeword $\bq_{\rm NF}(\phi, \tilde r_{1})$ is selected when
$
\frac{{\rm abs}\left( G\left(\bq_{\rm NF}(\phi, \tilde r_{1})\right) - G\left(\bq_{\rm NF}(\phi, \tilde r_{2})\right) \right) }{\max\left( G\left(\bq_{\rm NF}(\phi, \tilde r_{1})\right), G\left(\bq_{\rm NF}(\phi, \tilde r_{2})\right) \right)} \leq \epsilon_{\rm gain}
$ and
$
G\left(\bq_{\rm NF}(\phi, \tilde r_{1})\right) > G\left(\bq_{\rm FF}(\phi)\right)
$,
or when
$G\left(\bq_{\rm NF}(\phi, \tilde r_{1})\right) > G\left(\bq_{\rm NF}(\phi, \tilde r_{2})\right)$.
Here, $\epsilon_{\rm gain}$ denotes a predefined gain-difference threshold, which can be empirically selected and is set to $10^{-2}$ in this paper.
When $
G\left(\bq_{\rm NF}(\phi, \tilde r_{2})\right) > G\left(\bq_{\rm FF}(\phi)\right)
$ and $
G\left(\bq_{\rm NF}(\phi, \tilde r_{2})\right) > G\left(\bq_{\rm NF}(\phi, \tilde r_{1})\right)
$, the NF codeword $\bq_{\rm NF}(\phi, \tilde r_{2})$ is selected. In all other cases, the FF codeword $\bq_{\mathrm{FF}}(\phi)$ is selected.

The proposed mechanism enables adaptive NF-FF beam switching without prior knowledge of the user location or reliance on the Rayleigh distance. By selecting the beam according to the measured beamforming gains, it remains applicable during BT while introducing only limited additional overhead, as at most two NF codewords are compared with the FF codeword.

\subsection{Extension to Multi-Path Near-Field Beam Training}
In this section, we extend the proposed HSABT framework to multi-path channels by incorporating an adaptive candidate tracking strategy. Specifically, the multi-path channel of user is modeled as
\begin{equation}
	\begin{split}
		\bh_{\rm u} = \sum_{\ell=1}^{L_{\rm u}} \sqrt{N}  \beta_{{\rm u},\ell} \ba(r_{{\rm u},\ell},\theta_{{\rm u},\ell}),
	\end{split}
\end{equation}
where $L_{\rm u}$ denotes the number of propagation paths, and $\beta_{{\rm u},\ell}$, $\theta_{{\rm u},\ell}$, and $r_{{\rm u},\ell}$ represent the complex gain, angle, and distance of the $\ell$-th path, respectively.

Specifically, in the first stage, where a small-scale array is employed, the received beam gain profile is initially acquired in the angular domain. Since the predefined DFT beams provide only discrete angular measurements, cubic spline interpolation is adopted to reconstruct a continuous beam gain response. The dominant angular directions are subsequently identified by detecting the prominent peaks of the interpolated response. 
Accordingly, the number of candidate directions is adaptively determined based on the number of significant energy peaks, eliminating the need for prior knowledge of the number of propagation paths.
The identified angular candidates are subsequently refined in the polar domain, where multiple candidate angle-distance pairs associated with different propagation paths can be adaptively obtained.

Subsequently, each obtained candidate region is regarded as a parent region to generate two child regions at the next level. A child region is retained only when the ratio of its received gain to that of the parent region exceeds a threshold $\xi_a$.
As the hierarchical refinement proceeds, the NF effect becomes increasingly significant, leading to the coupling between the angular and distance domains. Despite this coupling, the energy consistency criterion is still employed for candidate validation. By leveraging the inherent energy consistency across successive hierarchical levels, the proposed method adaptively retains multiple candidate angles and distances while effectively suppressing spurious candidate regions.

\subsection{Hierarchical Sub-Array Codebook Design} 
In order to implement the proposed HSABT strategy, it is necessary to design a corresponding matched codebook.
Then, we first assume that the number of radiating elements in the $l$-th sub-array is $N_l$.
When $N_l \leq \tilde{N}$, the BT range lies in the FF region; otherwise, it lies in the NF range.
Accordingly, the design of the $l$-th sub-array codebook can be divided into two cases, i.e., $N_l \leq \tilde{N}$ and $N_l > \tilde{N}$.

\noindent $\blacklozenge$ {\bf{Case 1:}} When $N_l \leq \tilde{N}$, it is sufficient to perform sampling only in the angular domain, and the steering vector corresponding to each sampling point can be directly used as the associated codeword $\bw_{i}^l$ as follows
\begin{equation} \label{FF_Codeword}
	\begin{split}
		[\bq_{i}^l]_n = e^{j k_c \delta_n d \phi_i^l},~i,n=1,2,\dots,N_l,
	\end{split}
\end{equation}
where $\phi_i^l = \frac{2i-N_l-1}{N_l},~i=1,2,\dots,N_l$.

\noindent $\blacklozenge$ {\bf{Case 2:}} When $N_l > \tilde{N}$, sampling needs to be performed simultaneously in both the angular and distance domains, and the steering vector corresponding to each sampling point can be used as the associated codeword as follows
\begin{equation} \label{NF_Codeword}
	\begin{split}
		[\bq_{i,z}^l]_n = \sqrt{N_l}\cdot[\ba(r_{i,z}^l, \phi_i^l)]_n,~i,n=1,2,\dots,N_l,
	\end{split}
\end{equation}
where $\phi_i^l = \frac{2i-N_l-1}{N_l}$ and $r_{i,z}^l = \frac{1}{z}Z_{\Delta}(\phi_i^l),~z = 1,2,\dots,Z$.

It should be emphasized that the codewords in~\eqref{FF_Codeword} and~\eqref{NF_Codeword} are effective codewords, arising from the integration of FIM-based surface shaping and the discrete phase quantization of the steering vectors corresponding to the sampling points, which are subsequently mapped onto the discrete PS. 
Accordingly, during BT, each discrete PS must be strictly aligned with its designed FIM surface shape morphing.

Furthermore, the hierarchical codebooks of HSABT are generated offline and stored prior to transmission. Hence, its online cost is dominated by beam measurements rather than codebook construction. Let $M_{\rm HSABT}$ denote the training overhead and $T_{\rm P}$ denote the duration of one BT pilot. 
The corresponding online training latency is $T_{\rm HSABT}^{\rm online} = M_{\rm HSABT} T_{\rm P}$, which scales linearly with the training overhead. Since each beam measurement only involves received-gain evaluation, comparison, and candidate-region selection, the online processing complexity is also ${\cal O}(M_{\rm HSABT})$.

\vspace{-0.2cm}
\section{Multiuser Communications} \vspace{-0.05cm}
The proposed HSABT efficiently identifies the dominant propagation path of each user, thereby providing the channel information required for hybrid beamforming design in high-quality multi-user communications. In the considered hybrid architecture, the analog beamformer generates highly directional beam patterns to enhance the desired signal power, while the digital beamformer suppresses inter-user interference. Based on the acquired dominant-path information, we next develop a systematic design framework for the analog and digital beamformers.

\vspace{-0.3cm}
\subsection{Analog Beamformer Design}
In the context of multi-beam design, the surface shaping of the FIM is unable to completely suppress the grating-lobe effect.
However, by appropriately optimizing the surface shape, the gains of multiple beams can still be effectively enhanced. 
Accordingly, two different surface-shaping design strategies are developed, referred to as the alternating optimization (AO)-based approach and the key-observation-based approach.

\subsubsection{Alternating Optimization-Based Surface Shaping}
Our objective is to jointly optimize the surface shape and the analog beamformer to ensure that the beam gain at the user’s location closely matches the desired beam gain.
The corresponding optimization problem can be formulated as
\begin{equation} \label{original_prob}
	\begin{split}
		\min_{\Delta\by, \bV, \bg} &~ \sum_{k=1}^{K} \left| \ba_k(\Delta\by)^{\rm H}\bv_k - g_k \right|^2 \\
		{\rm s.t.}
		&~ -\frac{\hat y}{2} \leq \Delta y_n \leq \frac{\hat y}{2},~|g_k|=N, \\
		&~ [ \bv_k ]_n = e^{j \tau_{k,n}},~\tau_{k,n} \in {\cal C}_b,~k=1,\dots,K,
	\end{split}
\end{equation}
where $\ba_k(\Delta\by) = \left[e^{j \hat\varphi_k(\Delta y_1)}, e^{j \hat\varphi_k(\Delta y_2)}, \dots, e^{j \hat\varphi_k(\Delta y_N)} \right]^{\rm T}$ and $\hat\varphi_k(\Delta y_n) = \varphi_n(\Delta y_n)$ with respect to $\theta_k$.

Then, we develop an AO method to solve the problem~\eqref{original_prob}.

\noindent $\blacklozenge$ {\bf{Optimize $\bg$:}}
Given $\Delta\by$ and $\bV$, the optimal solution is
\begin{equation}
	\begin{split}
		g_k = N\cdot e^{j \angle(\ba_k^{\rm H}\bv_k)}.
	\end{split}
\end{equation}

\noindent $\blacklozenge$ {\bf{Optimize $\bv_k$:}}
Given $\Delta\by$ and $\bg$, the problem~\eqref{original_prob} is equivalent to solving the following $K$ problems in parallel.
\begin{equation} \label{prob_u_k}
	\begin{split}
		\min_{\bv_k} &~ \left| \ba_k^{\rm H}(\Delta\by)\bv_k - g_k \right|^2 \\
		{\rm s.t.}
		&~ [ \bv_k ]_n = e^{j \tau_{k,n}},~\tau_{k,n} \in {\cal C}_b.
	\end{split}
\end{equation}

For the sake of convenience, we abbreviate $\ba_k(\Delta\by)$ as $\ba$, $\bv_k$ as $\bv$, and $g_k$ as $g$.
We then use the proximal distance algorithm (PDA)~\cite{Zhang2024PDA,Zhang2025FA_ISAC} to transform the problem~\eqref{prob_u_k} into
\begin{equation} \label{prob_u_k_PDA}
	\begin{split}
		\min_{\bv} ~ \left| \ba^{\rm H}\bv - g \right|^2 + \mu \left\| \bv - Q_b(\hat\bv) \right\|_2^2,
	\end{split}
\end{equation}
where $\mu > 0$ is a penalty factor, $\hat\bv$ is the optimal solution in the previous step.
We can derive the optimal closed-form solution as
\begin{equation}
	\begin{split}
		\bv^\star = \left( \frac{1}{\mu}\bI - \frac{\ba \ba^{\rm H}}{\mu(\mu + \|\ba \|_2^2) } \right) \left( g \ba + \mu Q_b(\hat\bv) \right).
	\end{split}
\end{equation}

\noindent $\blacklozenge$ {\bf{Optimize $\Delta\by$:}}
To optimize $\Delta\by$, we first develop the following variables substitutions based on hyperbolic tangent function $\rm tanh(\cdot)$, given by
\[
\Delta y_n = \frac{-\frac{\hat y}{2} + \frac{\hat y}{2}}{2} + \frac{\frac{\hat y}{2} + \frac{\hat y}{2}}{2} \tanh(\alpha z_n) = \frac{\hat y}{2} \tanh(\alpha z_n),~z_n\in \mathbb{Z}.
\]
\[
\frac{\partial \Delta y_n}{\partial z_n} = \frac{\alpha\hat y}{2}\left( 1 - \tanh^2(\alpha z_n) \right),
\]
where $\alpha > 0$ is a given parameter.

We then define $f_k(\Delta\by) = \ba(\Delta\by)^{\rm H} \bv_k$ and then have
\[
f_k(\bz) = \sum_{n=1}^{N} v_{k,n} e^{-j \hat\varphi_k(z_n)},
\]
where $\hat\varphi_k(z_n) = k_c \left( \delta_n d \cdot \sin\theta_k + \hat y \tanh(\alpha z_n) \cos\theta_k / 2 \right) - k_c\left( \hat y \tanh(\alpha z_n) \sin\theta_k / 2 - \delta_n d \cdot \cos\theta_k \right)^2\big/ 2r_k.$

Then, the problem~\eqref{original_prob} with respect to $\Delta\by$ can be transformed into the {\it residual minimization} problem, given by
\begin{equation} \label{prob_z}
	\begin{split}
		\min_{\bz} &~ {\cal F}(\bz) = \sum_{k=1}^{K} \left| \epsilon_k(\bz) \right|^2,
	\end{split}
\end{equation}
where $\epsilon_k(\bz) = f_k(\bz) - g_k = \sum_{n=1}^{N} v_{k,n} e^{-j \hat\varphi_k(z_n)} - g_k$.

Further, we can rewrite the problem~\eqref{prob_z} as
\begin{equation} \label{prob_z_R_I}
	\begin{split}
		\min_{\bz} &~ \sum_{k=1}^{K} \left( \Re\{\epsilon_k(\bz)\} \right)^2 + \left( \Im\{\epsilon_k(\bz)\} \right)^2,
	\end{split}
\end{equation}
where
$$
\Re\{\epsilon_k(\bz)\} = \sum_{n=1}^{N} | v_{k,n} | \cos\left( \hat\varphi_k(z_n) - \angle(v_{k,n}) \right) - \Re\{g_k\},
$$
$$
\Im\{\epsilon_k(\bz)\} = \sum_{n=1}^{N} - | v_{k,n} | \sin\left( \hat\varphi_k(z_n) - \angle(v_{k,n}) \right) - \Im\{g_k\}.
$$

The problem can be solve by the gradient descent algorithm~\cite{Zhang2025FA_ISAC,Zhang2024Practical_RIS}, which follows
\begin{equation}
	\begin{split}
		\bz^{\ell+1} = \bz^{\ell} - \eta\cdot \frac{\partial {\cal F}(\bz)}{\partial \bz},
	\end{split}
\end{equation}
where $\eta > 0$ is a step length that can be found by the backtracking line search~\cite{Boyd2004convex} and
\begin{equation*}
	\begin{split}
		&\frac{\partial {\cal F}(\bz)}{\partial z_n} =  \sum_{k=1}^{K} 2 \Re\{ \epsilon_k(\bz) \}\left( - \sin\left( \varphi_k(z_n) - \angle(v_{k,n}) \right) \frac{\partial \hat\varphi_k(z_n)}{\partial z_n} \right) \\
		&- \sum_{k=1}^{K} 2 \Im\{ \epsilon_k(\bz) \}\left( \cos\left( \varphi_k(z_n) - \angle(v_{k,n}) \right) \frac{\partial \hat\varphi_k(z_n)}{\partial z_n} \right),
	\end{split}
\end{equation*}
where 
\[
\frac{\partial \hat\varphi_k(z_n)}{\partial z_n} = \frac{\partial \hat\varphi_k(z_n)}{\partial \Delta y_n}  \frac{\partial \Delta y_n}{\partial z_n} = \frac{\partial \hat\varphi_k(z_n)}{\partial \Delta y_n} \frac{\alpha\hat y\left( 1 - \tanh^2(\alpha z_n) \right)}{2},
\]
\begin{equation*}
	\begin{split}
		\frac{\partial \hat\varphi_k(z_n)}{\partial \Delta y_n} = &\frac{k_c}{r_k} \left( \delta_n d \cos\theta_k - \frac{\hat y}{2} \tanh(\alpha z_n) \sin\theta_k \right) \sin\theta_k\\
		& + k_c \cos\theta_k.
	\end{split}
\end{equation*}

{\it Remark 4 (Discussion on algorithm selection):}
The transformation $\Delta y_n = \frac{\hat y}{2} \tanh(\alpha z_n)$ guarantees deformation feasibility throughout the iterations while preserving the smooth differentiability of the coupled objective.
More importantly, through the chain rule, it introduces a state-dependent scaling of the physical displacement gradient and gradually reduces the update magnitude as an element approaches its deformation boundary. This implicit scaling and smooth boundary damping are particularly beneficial for the considered wavelength-scale deformation problem, whose phase sensitivities and local curvatures vary substantially across elements and users. In contrast, unconstrained gradient descent cannot directly enforce the deformation bounds, while projected gradient descent is more sensitive to variable scaling and step-size selection and may suffer from abrupt clipping and active-set switching at the boundaries. Manifold optimization is not directly applicable to the box-constrained deformation variables or the finite discrete phase set, and continuous phase relaxation followed by quantization may introduce a relaxation-quantization mismatch. Successive convex approximation is applicable in principle, but repeatedly constructing and solving surrogate subproblems for the strongly coupled complex-exponential objective incurs higher computational and memory costs. Therefore, the $\rm tanh$-based gradient algorithm provides a favorable balance among feasibility, numerical stability, implementation simplicity, and per-iteration complexity.

\subsubsection{Key-Observation-Based Surface Shaping}
In this section, we aim to develop a non-iterative approach capable of directly obtaining the FIM surface shape. 
Inspired by the FIM compensation-based quantization error method introduced in Section III, we propose to shape the FIM surface such that the phase quantization errors across all user steering vectors are minimized, a strategy we refer to as the key observation method (KOM).
Based on this, the surface shape of the FIM can be obtained by simultaneously solving $N$ problems as follows
\begin{equation} \label{error_compensation_multi_beam}
	\begin{split}
		\min_{ \Delta y_n \in [-\frac{\hat y}{2}, \frac{\hat y}{2}] }~ \sum_{k=1}^{K}\left(A_{n,k} \Delta y_n^2 + B_{n,k} \Delta y_n + C_{n,k} \right)^2,
	\end{split}
\end{equation}
where 
$$
A_{n,k} = \frac{-k_c \sin^2\theta_k}{2 r_k},
B_{n,k} = \frac{k_c\cos\theta_k( \delta_n d  \sin\theta_k + r_k)}{r_k}.$$
When $\widetilde\varphi_k(0) \leq Q_b\left( \widetilde\varphi_k(0) \right)$, $C_{n,k} = - \left| \widetilde\varphi_k(0) - Q_b\left( \widetilde\varphi_k(0) \right) \right| $; otherwise, $C_{n,k} = \left| \widetilde\varphi_k(0) - Q_b\left( \widetilde\varphi_k(0) \right) \right| $, where $\widetilde\varphi_k(0) = {\rm mod}(\varphi_k(0),2\pi)$.

Then, the problem ~\eqref{error_compensation_multi_beam} can be reformulated as follows
\begin{equation*} \label{four_square}
	\begin{split}
		\min_{ \Delta y_n \in [-\frac{\hat y}{2}, \frac{\hat y}{2}] }~ {\cal G} = \alpha_4 \Delta y_n^4 + \alpha_3 \Delta y_n^3 + \alpha_2 \Delta y_n^2 + \alpha_1 \Delta y_n + \alpha_0,
	\end{split}
\end{equation*}
where $\alpha_{0} = \sum_{k=1}^{K} C_{n,k}^2$, $\alpha_1 = 2\sum_{k=1}^{K}B_{n,k}C_{n,k} $,  $\alpha_2 = \sum_{k=1}^{K}\left(B_{n,k}^2 + 2A_{n,k}C_{n,k} \right)$, $\alpha_3 = 2\sum_{k=1}^{K} A_{n,k} B_{n,k}$, and $\alpha_4 = \sum_{k=1}^{K} A_{n,k}^2 $.

Subsequently, we derive the derivative of the objective function with respect to $\Delta y_n$.
\begin{equation}
	\begin{split}
		\frac{\partial {\cal G}}{\partial \Delta y_n} = 4\alpha_4 \Delta y_n^3 + 3\alpha_3 \Delta y_n^2 + 2\alpha_2 \Delta y_n + \alpha_1.
	\end{split}
\end{equation}

The optimal $\Delta y_n$ is determined by evaluating the roots of $\frac{\partial {\cal G}}{\partial \Delta y_n} = 0$ within the constraint set and the boundary points, and choosing the value that minimizes the objective function.

\vspace{-0.3cm}
\subsection{Digital Beamformer Design} \vspace{-0.1cm}
We first define the equivalent channel of user $k$ as $\widetilde\bh_k^{\rm H} = \bh_k^{\rm H} \bV$. 
Our objective is to leverage the digital beamformer $\bW$ to manage multiuser interference so as to maximize the multiuser sum rate.
Therefore, the multiuser sum rate maximization problem is given by
\begin{equation} \label{SRM_prob}
	\begin{split}
		\max_{\bW}&~ \sum_{k=1}^{K} {\rm log}_2\, \left( 1 + \frac{ \left| \widetilde\bh_k^{\rm H} \bw_k \right|^2 }{\sum_{i=1,i\neq k}^{K} \left| \widetilde\bh_k^{\rm H} \bw_i \right|^2 + \sigma_k^2 } \right) \\
		 {\rm s.t.}&~\sum_{k=1}^{K} \left\| \bV \bw_k \right\|_2^2 \leq P_{\rm max}.
	\end{split}
\end{equation}

By the weighted minimum-mean-square-error (WMMSE) algorithm~\cite{Shi2011WMMSE,Zhang2024PDA}, the problem~\eqref{SRM_prob} can be transformed into
\begin{equation} \label{SRM_prob_WMMSE}
	\begin{split}
		\min_{\bW} &~ \sum_{k=1}^{K} \left[ \bw_k^{\rm H} \bE \bw_k - 2 \Re\{ \be_k^{\rm H}\bw_k \} \right] \\
		{\rm s.t.}&~\sum_{k=1}^{K} \left\| \bV \bw_k \right\|_2^2 \leq P_{\rm max},
	\end{split}
\end{equation}
where $\bE = \sum_{k=1}^{K} \rho_k |u_k|^2 \widetilde\bh_k \widetilde\bh_k^{\rm H}$, $\be_k = \rho_k u_k \widetilde\bh_k$, and 
\begin{equation*}
	\begin{split}
		u_k = \frac{\widetilde\bh_k^{\rm H} \hat\bw_k}{\sum_{i=1}^{K} \left| \widetilde\bh_k^{\rm H} \hat\bw_i \right|^2 + \sigma_k^2 },~\rho_k = \frac{1}{ 1 - u_k^* \widetilde\bh_k^{\rm H} \hat\bw_k},
	\end{split}
\end{equation*}
where $\hat\bw_k$ is the optimal solution in the previous iteration.

Therefore, we can obtain the closed-form solution of the problem~\eqref{SRM_prob_WMMSE}, which is given by
\begin{equation}
	\begin{split}
		\bw_k^\star = \left(\bE + \zeta \bV^{\rm H}\bV  \right)^{-1} \be_k,~k=1,2,\dots,K,
	\end{split}
\end{equation}
where $\zeta$ is a dual variable that can be found by the one-dimensional search.

The AO-based design has complexity ${\cal O}(I_{AO} NK[1+I_P+I_G(1+I_{LS})])$, whereas the non-iterative KOM has complexity $\mathcal{O}(NK)$.
The complexity of WMMSE algorithm is ${\cal O}( NN_{RF}K + NN_{RF}^2 + I_W[ N_{RF}K^2 + KN_{RF}^2 + I_{\zeta}(N_{RF}^3 + N_{RF}^2K) ] )$.
A comprehensive theoretical analysis of the algorithms developed in this section, including their convergence behavior, the characterization of the resulting block-wise steady points, and their computational complexity, is provided in the supplementary document available at \url{https://github.com/qianzhangcn/HSABT_FIM_Supplementary_Document}.

\section{Simulation Results}
In this section, simulation results are provided to validate the accuracy and performance of the proposed HSABT scheme, as well as to demonstrate the effectiveness of the FIM-enhanced multiuser communications. 
The carrier frequency is set to 30 GHz, with $512$ radiating elements and a surface shaping range of $\hat y = 2\lambda$. 

\begin{figure}[t]
	\centering
	\vspace{-0.25cm}
	\subfloat[1-st level $N_1=16$]{\includegraphics[width=0.48\columnwidth]{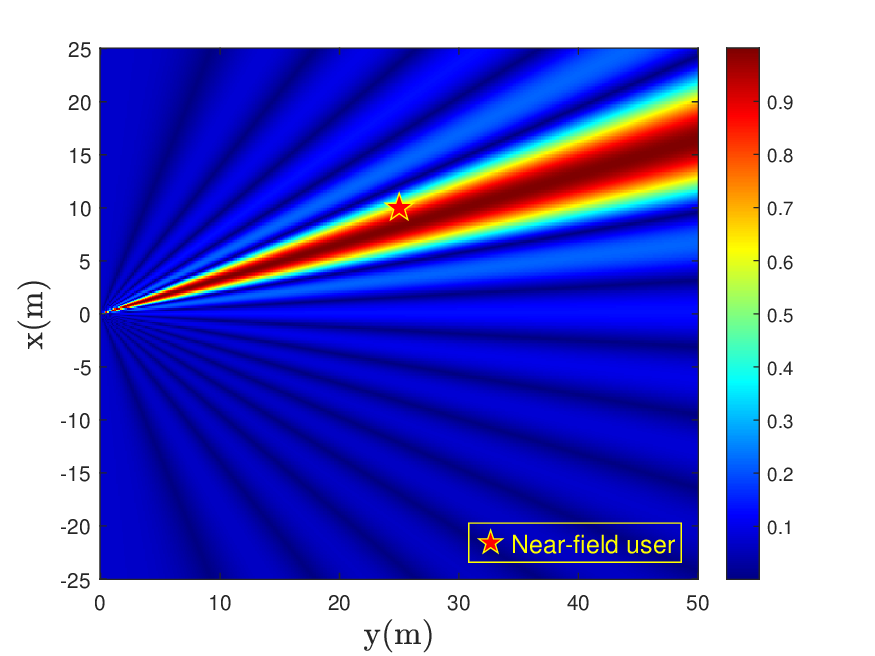}%
		\label{RHBT_BeamPattern_01}}
	\hfil
	\subfloat[2-nd level $N_2=32$]{\includegraphics[width=0.48\columnwidth]{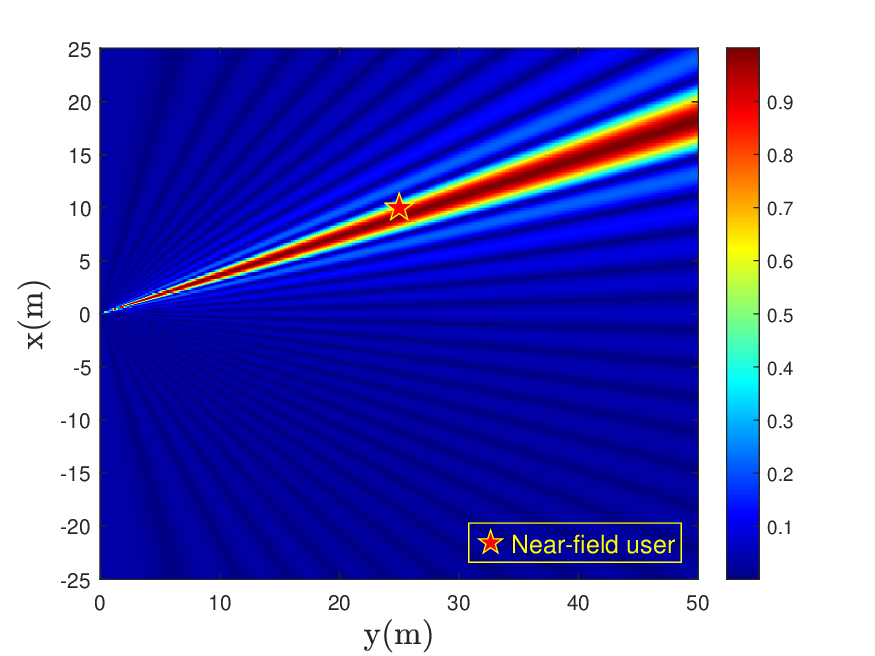}%
		\label{RHBT_BeamPattern_02}}
	\hfil
	\vspace{-0.2cm}
	\subfloat[3-rd level $N_3=64$]{\includegraphics[width=0.48\columnwidth]{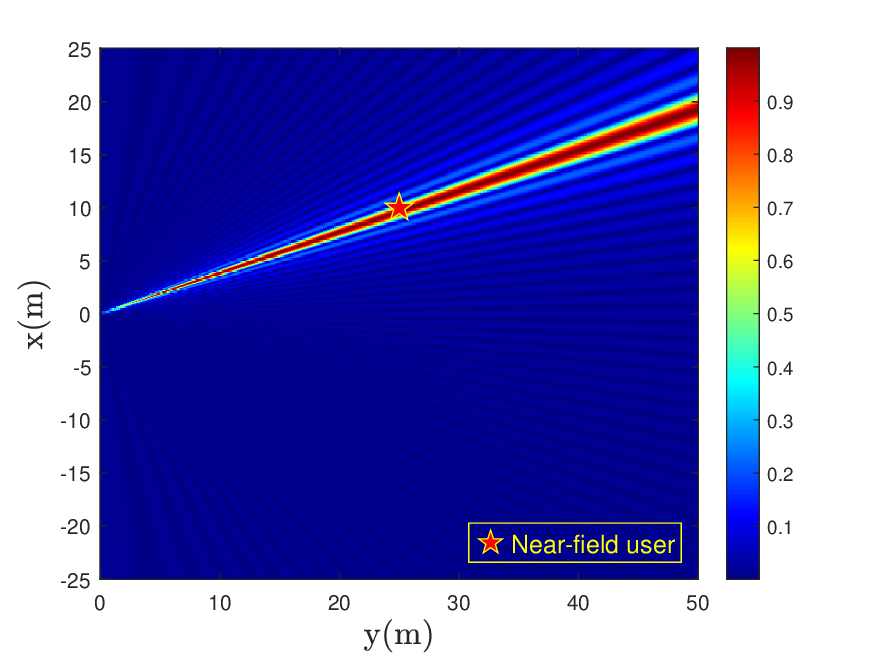}%
		\label{RHBT_BeamPattern_03}}
	\hfil
	\subfloat[4-th level $N_4=128$]{\includegraphics[width=0.48\columnwidth]{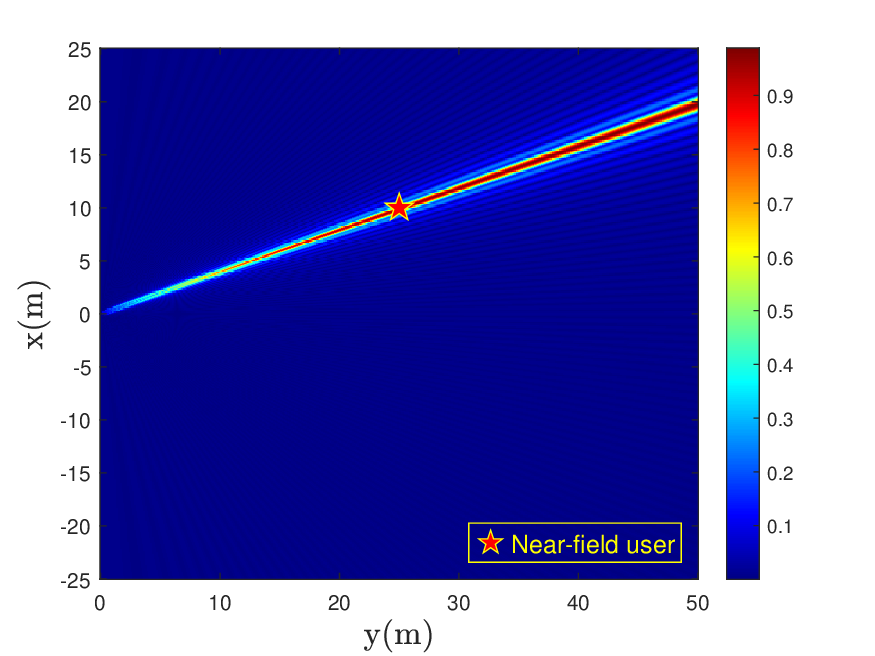}%
		\label{RHBT_BeamPattern_04}}
	\hfil
	\vspace{-0.2cm}
	\subfloat[5-th level $N_5=256$]{\includegraphics[width=0.48\columnwidth]{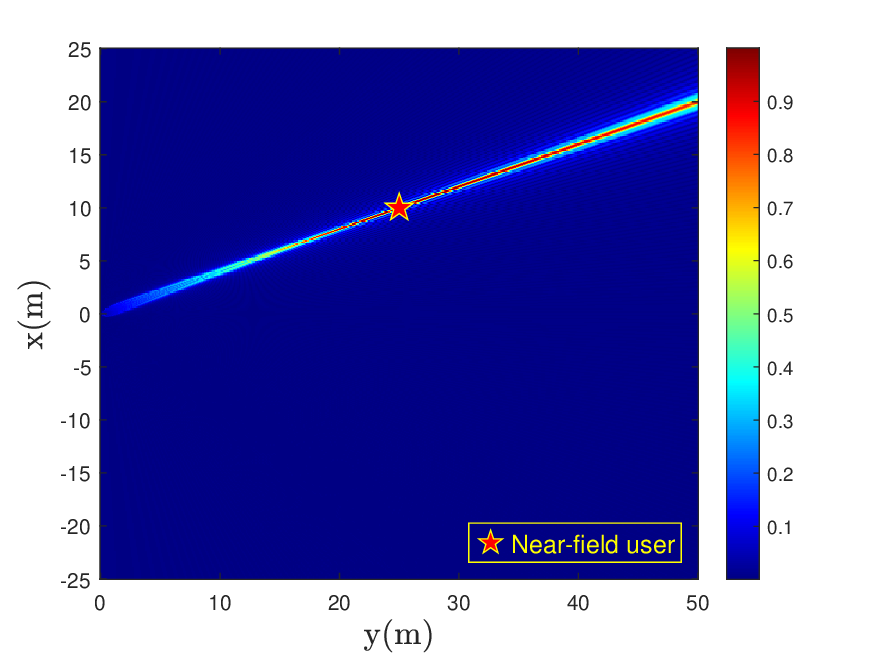}%
		\label{RHBT_BeamPattern_05}}
	\hfil
	\subfloat[6-th level $N_6=512$]{\includegraphics[width=0.48\columnwidth]{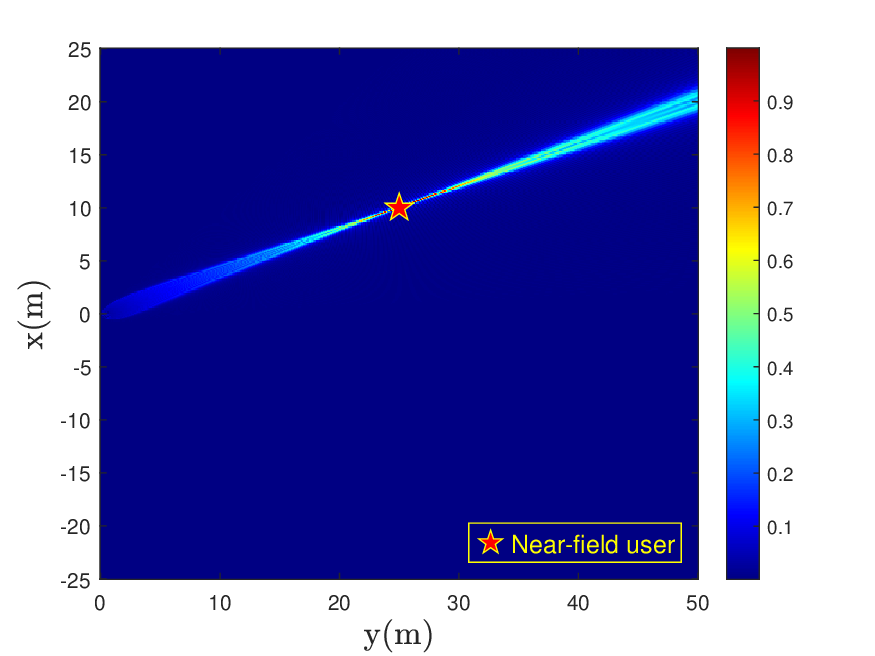}%
		\label{RHBT_BeamPattern_06}}
	\caption{The near-field beam pattern under different level beam training.}
	\vspace{-0.45cm}
	\label{fig:RHBT_BeamPattern}
\end{figure}
\begin{figure}[t]	
	\centering \includegraphics[width=0.8\linewidth]{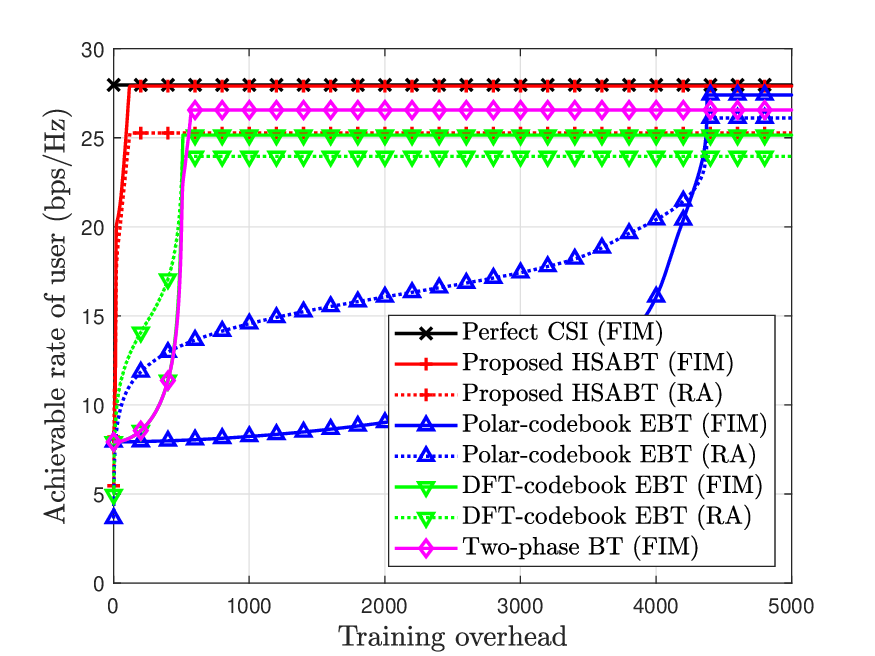}
	\vspace{-0.3cm}
	\caption{Achievable rate versus training overhead under different beam training schemes, SNR=30$\,$dB, $N=512$.}
	\vspace{-0.5cm}
	\label{fig:Rate_overhead}
\end{figure}
\begin{figure}[t]	
	\centering \includegraphics[width=0.8\linewidth]{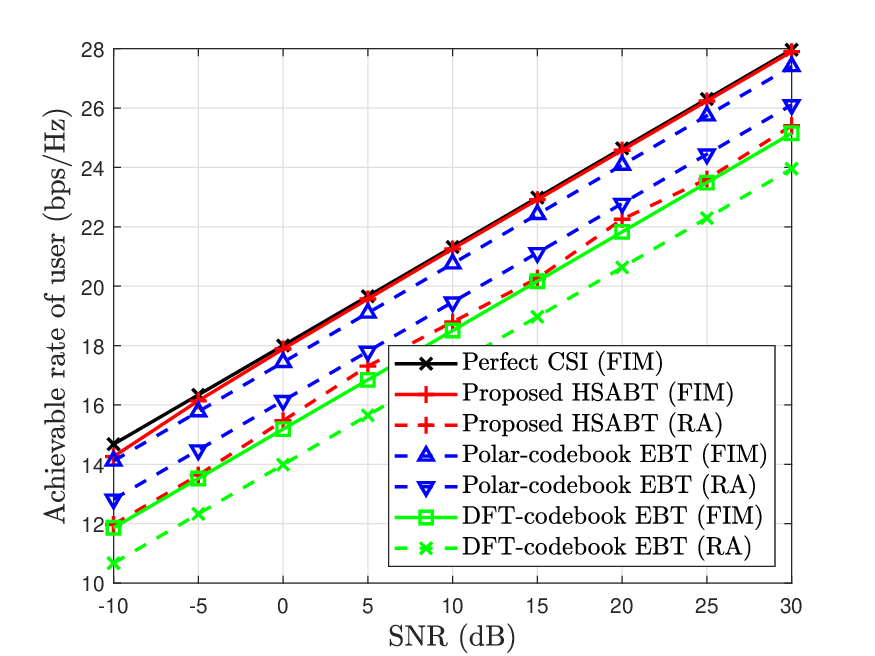}
	\vspace{-0.3cm}
	\caption{Achievable rate versus SNR under different beam training schemes with the assistance of FIM, $N=512$.}
	\vspace{-0.5cm}
	\label{fig:Rate_SNR}
\end{figure}
\begin{figure}[t]	
	\centering \includegraphics[width=0.8\linewidth]{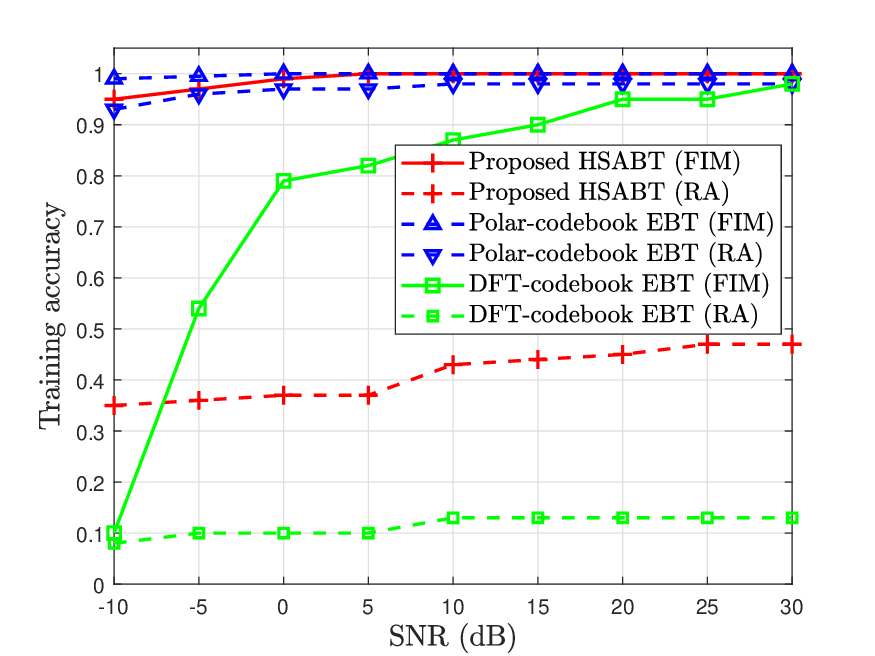}
	\vspace{-0.3cm}
	\caption{Training accuracy versus SNR under different beam training schemes.}
	\vspace{-0.5cm}
	\label{fig:Training_accuracy_SNR}
\end{figure}
\begin{figure}[t]	
	\centering \includegraphics[width=0.8\linewidth]{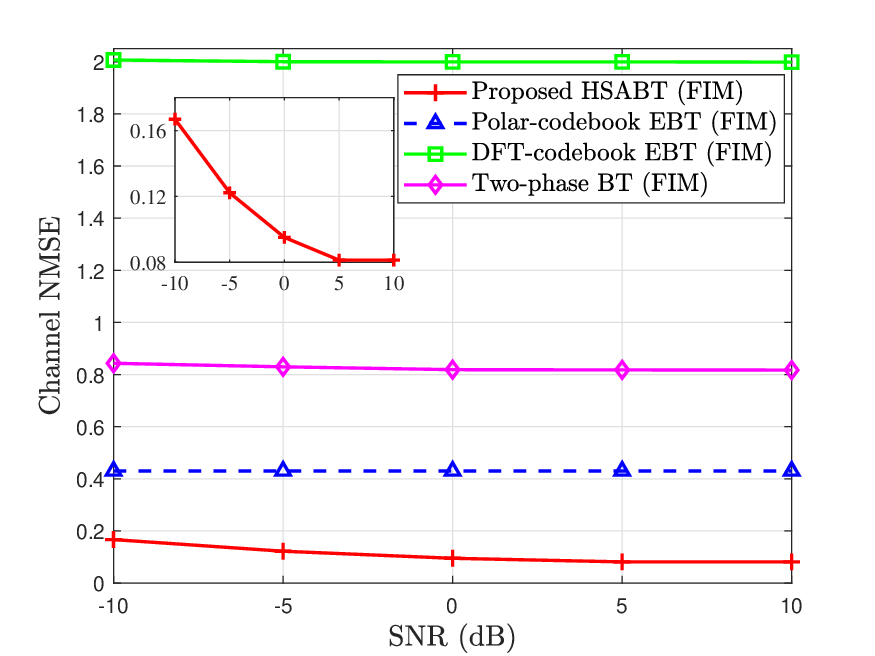}
	\vspace{-0.3cm}
	\caption{Comparison of channel normalized mean square error (NMSE) obtained by different beam training methods, $N=512$.}
	\vspace{-0.45cm}
	\label{fig:Channel_NMSE}
\end{figure}
\begin{figure}[t]	
	\centering \includegraphics[width=0.8\linewidth]{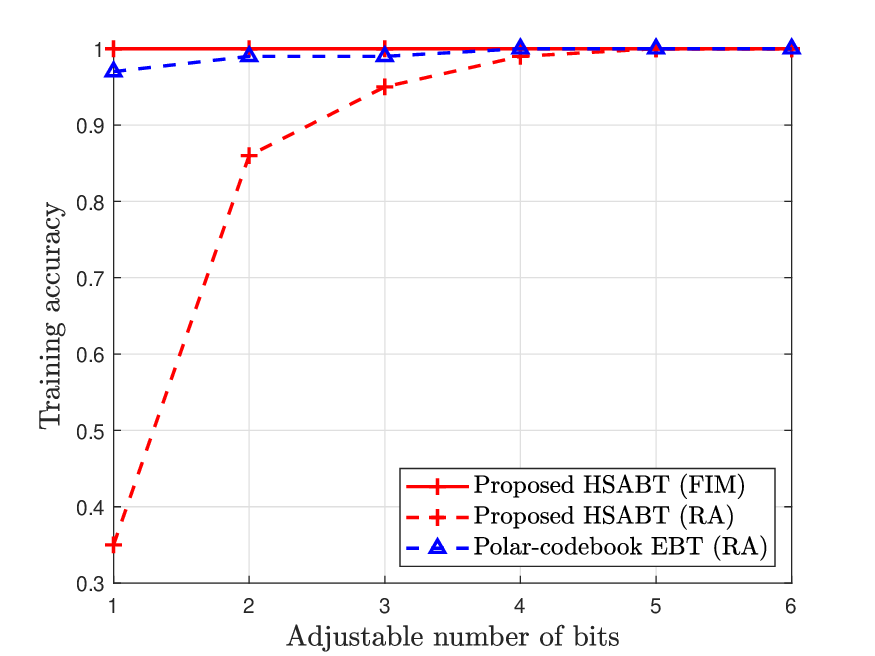}
	\vspace{-0.3cm}
	\caption{Beam-training accuracy of the considered schemes under different PSs quantization resolutions.}
	\vspace{-0.25cm}
	\label{fig:Training_accuracy_bits}
\end{figure}
\begin{figure}[t]
	\centering
	\vspace{-0.25cm}
	\subfloat[True paths, $\beta_{{\rm u},\ell}=\{0,-12,-6\}\,$dB]{\includegraphics[width=0.5\columnwidth]{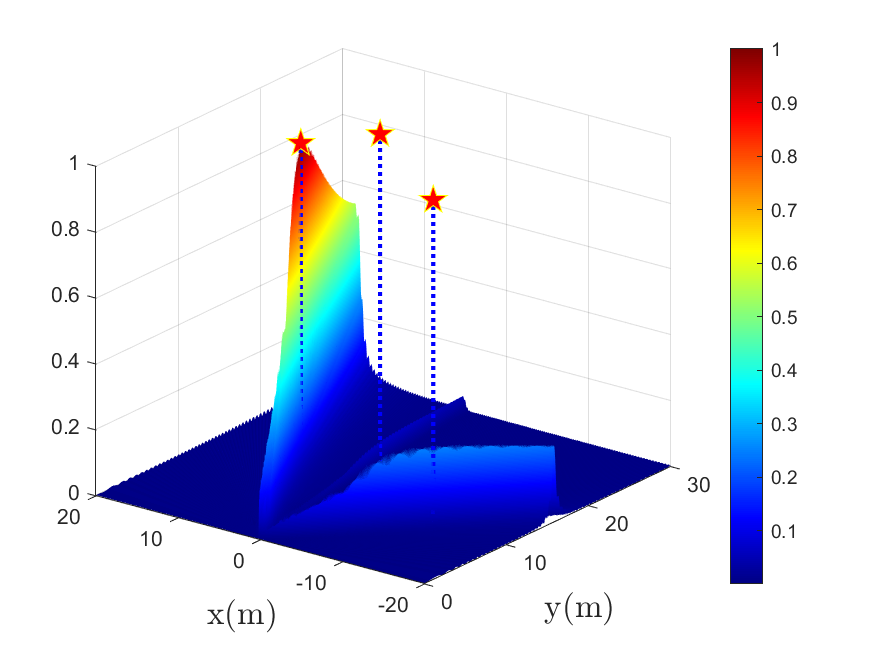}%
		\label{User_0__6__3}}
	\hfil
	\subfloat[Multi-path BT result]{\includegraphics[width=0.5\columnwidth]{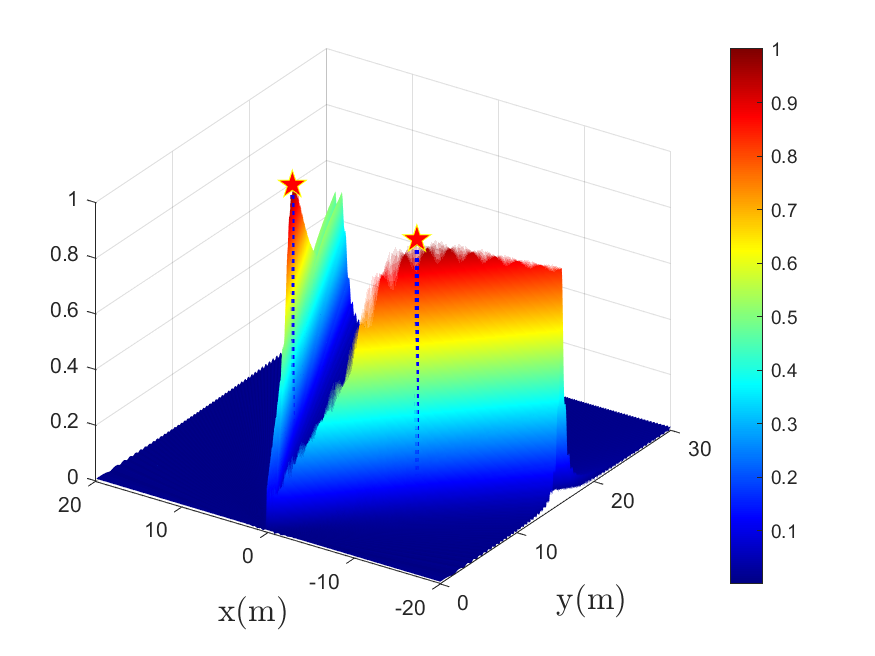}%
		\label{Multipath_BT_0__6__3}}
	\hfil
	\vspace{-0.2cm}
	\subfloat[True paths, $\beta_{{\rm u},\ell}=\{0,0,0\}\,$dB]{\includegraphics[width=0.5\columnwidth]{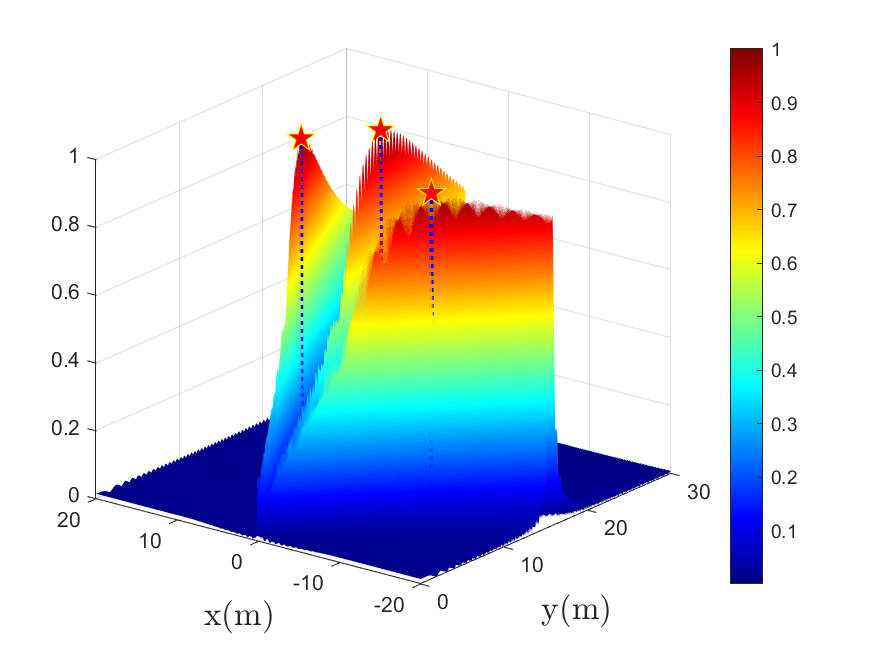}%
		\label{User_0_0_0}}
	\hfil
	\subfloat[Multi-path BT result]{\includegraphics[width=0.5\columnwidth]{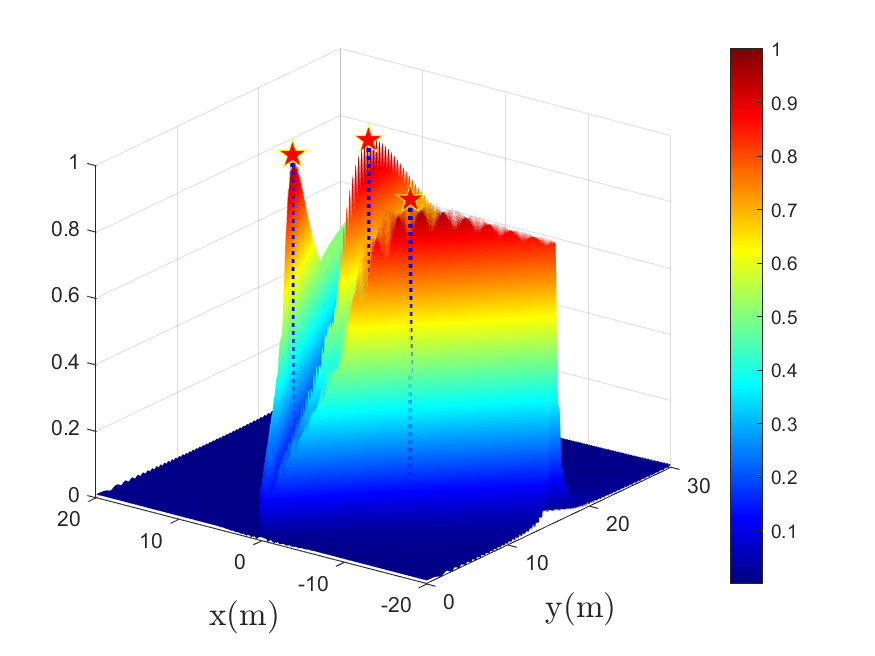}%
		\label{Multipath_BT_0_0_0}}
	\caption{Multi-path BT results with FIM for $L_{\rm u}=3$. Stars denote the true paths in the left column and the paths detected by HSABT in the right column.}
	\vspace{-0.5cm}
	\label{fig:Multipath_BT}
\end{figure}
\begin{figure}[t]	
	\centering \includegraphics[width=0.85\linewidth]{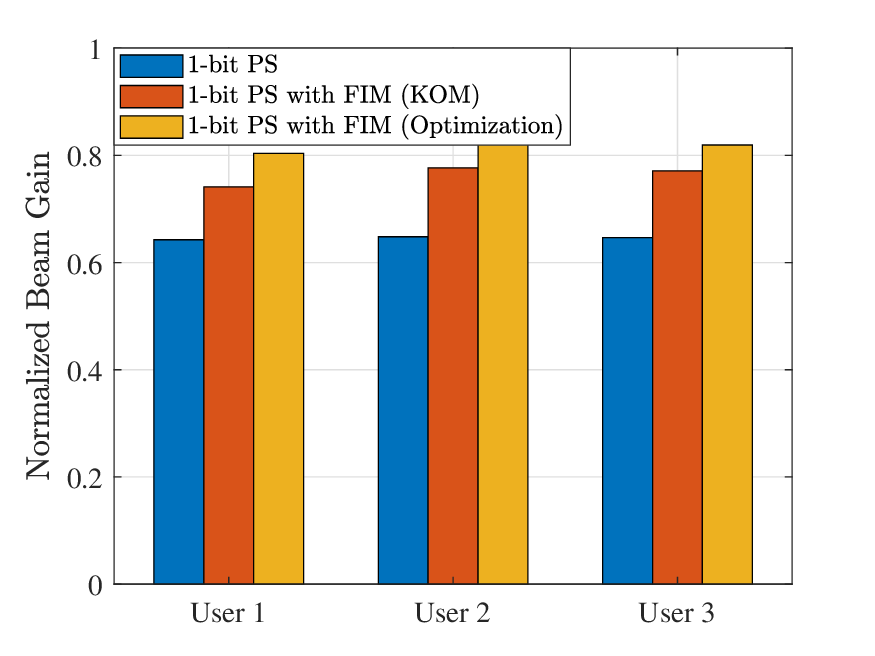}
	\vspace{-0.3cm}
	\caption{The near-field normalized beam gain for multi-beam scenarios under different phase shifters.}
	\vspace{-0.45cm}
	\label{fig:Multi_Beam_bar}
\end{figure}

\vspace{-0.3cm}
\subsection{Hierarchical Sub-Array Beam Training}
In our design, the number of radiating elements in the first-level sub-array is set to $N_1 = 16$. Consequently, the array can be hierarchically divided into $1 + {\rm log}_2(N/N_1)$ sub-array levels. 
As illustrated in Fig.~\ref{fig:RHBT_BeamPattern}, the proposed HSABT scheme generates six hierarchical beam patterns. 
It can be observed that each beam at every level precisely covers the user, and as the sub-array size increases progressively, the beam can accurately localize the user’s position.
To further demonstrate the superiority of the proposed HSABT scheme, we compare it with several benchmark methods in terms of both achievable rate and BT overhead. 
The benchmark schemes include DFT codebook-based BT, polar codebook-based BT, and two-phase BT~\cite{Zhang2022TPBT}. 
As shown in Fig.~\ref{fig:Rate_overhead}, we simulate the variation of the achievable rate with respect to the BT overhead.
Fig.~\ref{fig:Rate_overhead} shows that the proposed HSABT scheme achieves a higher achievable rate than exhaustive polar codebook training in capturing the dominant channel path, while requiring drastically lower overhead.
This is because the proposed HSABT realizes NF EBT for the entire array by performing BT at each sub-array level, as can also be observed from Fig.~\ref{fig:RHBT_BeamPattern}. 
Compared with the DFT-based and two-phase BT schemes, HSABT achieves a significantly higher achievable rate with approximately {\it 10 times} reduction in training overhead. 
Furthermore, Fig.~\ref{fig:Rate_overhead} demonstrates that BT performance is substantially improved when the FIM-based quantization error compensation is applied, outperforming the conventional RA-based BT.
As shown in Fig.~\ref{fig:Rate_SNR}, the proposed HSABT scheme consistently maintains superior performance at different signal-to-noise ratio (SNR) of $[-10,30]\,$dB.

As illustrated in Fig.\ref{fig:Training_accuracy_SNR}, the DFT codebook suffers from angular-domain spreading in NF scenarios. 
To achieve more accurate BT, we select the central angles corresponding to the five beams with the highest gains as the estimated user angles. 
It can be observed that when the SNR exceeds 0 dB, the proposed HSABT achieves the same training accuracy as the polar codebook while clearly outperforming the other benchmark methods. 
Moreover, by incorporating FIM-based quantization error compensation, the training accuracy is further and sharply improved.
Therefore, applying the proposed FIM-based quantization error compensation in future NF communication systems can substantially enhance both CSI estimation accuracy and overall communication performance.
Fig.~\ref{fig:Channel_NMSE} compares the channel NMSE achieved by different beam-training schemes versus the SNR. The proposed FIM-based HSABT consistently attains the lowest NMSE over the entire SNR range of $[-10,10]\,$dB, decreasing from approximately $0.16$ at $-10$ dB to below $0.1$ at moderate and high SNR. In contrast, the polar-codebook, two-phase, and DFT-codebook schemes exhibit pronounced error floors. In particular, the angular-only DFT codebook cannot accurately characterize the angle-distance coupling of NF channels, resulting in the largest estimation error. These results demonstrate that the proposed hierarchical refinement and beam-intersection-based BT jointly provide more accurate channel acquisition than the benchmark schemes.

Fig.~\ref{fig:Training_accuracy_bits} evaluates the BT accuracy of the considered schemes under different PSs resolutions. The proposed FIM-based HSABT maintains nearly perfect training accuracy over the entire range of quantization bits. In particular, its accuracy remains close to unity even with 1-bit PSs, demonstrating strong robustness against severe phase quantization errors. By contrast, the RA-based HSABT achieves an accuracy of only approximately 0.35 with 1-bit PSs, which increases progressively as the phase resolution improves and approaches unity when 4 or more quantization bits are employed. 
The polar codebook EBT with the RA also maintains high training accuracy under different PSs resolutions, but generally requires a substantially larger. These results demonstrate that wavelength-scale FIM morphing can effectively compensate for the phase errors introduced by low-resolution PSs, thereby preserving the desired hierarchical beam patterns and achieving reliable BT with low-resolution hardware.
Fig.~\ref{fig:Multipath_BT} evaluates the robustness of the proposed HSABT scheme under multi-path propagation and different path-power distributions. 
For $L_{\rm u}=3$ and $\beta_{{\rm u},\ell}=\{0,-12,-6\}\,$dB, the proposed method retains the candidate angle-distance regions associated with the dominant and relatively strong multi-path components, while the branch corresponding to the severely attenuated component is terminated by the peak-significance and inter-level energy-consistency criteria. 
This avoids unnecessary refinement of weak or spurious candidates and reduces the training overhead. 
When the three paths have equal gains, i.e., $\beta_{{\rm u},\ell}=\{0,0,0\}\,$dB, three distinct candidate regions are preserved throughout the hierarchical refinement. 
These results demonstrate that HSABT can adaptively adjust the number of retained candidates according to the observed energy distribution and provide reliable candidate directions and distances under multi-path propagation, shadowing, and partial blockage, without requiring prior knowledge of the number of significant paths.

\begin{figure}[t]	
	\centering \includegraphics[width=0.85\linewidth]{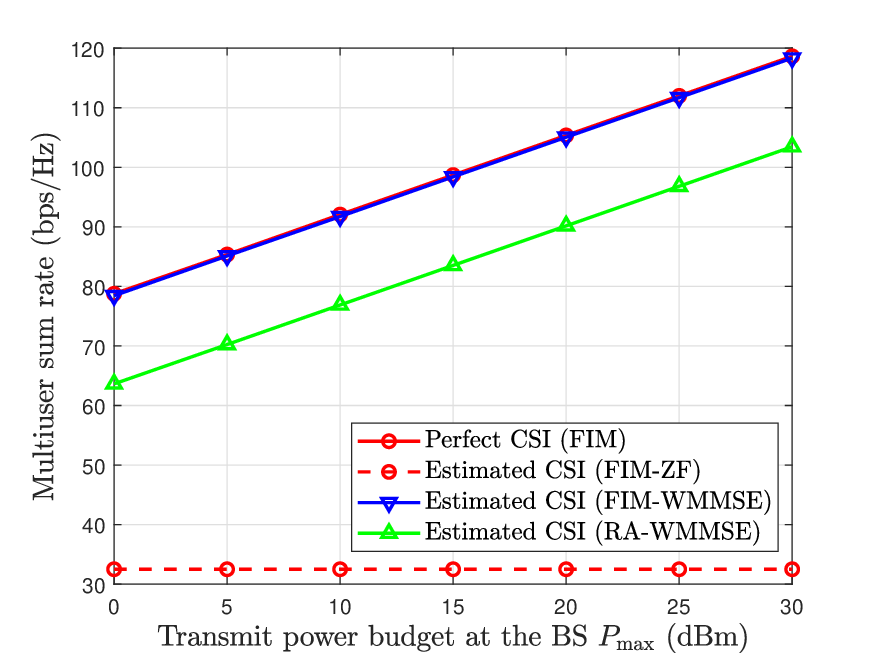}
	\vspace{-0.2cm}
	\caption{Multiuser sum rate performance under different transmit power budgets and CSI conditions, $N=512,K=4$.}
	\vspace{-0.45cm}
	\label{fig:SRM_P_max}
\end{figure}
\begin{figure}[t]	
	\centering \includegraphics[width=0.85\linewidth]{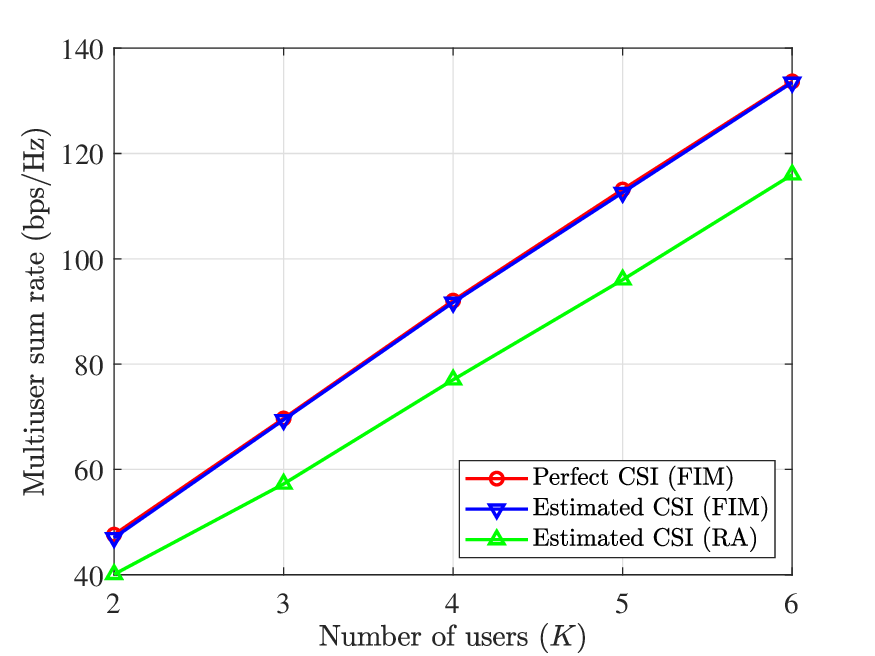}
	\vspace{-0.2cm}
	\caption{Multiuser sum rate performance under different number of users and CSI conditions, $N=512,P_{\rm max}=30\,$dBm.}
	\vspace{-0.3cm}
	\label{fig:SRM_K}
\end{figure}

\subsection{Multiuser Communications}
Based on the proposed HSABT scheme, accurate CSI for each user can be effectively obtained, enabling the joint optimization of the FIM surface shape and hybrid beamformers to enhance multiuser communication performance. 
As illustrated in Fig.~\ref{fig:Multi_Beam_bar}, optimizing the FIM surface shape and analog beamformer yields a substantial improvement in the beam gain of each user compared with conventional RA-based systems. 
Moreover, the proposed closed-form surface shaping design based on the KOM approach significantly reduces computational complexity compared with the optimization-based surface shaping method, while achieving a noticeable beam gain improvement over the conventional RA system.

Furthermore, Fig.~\ref{fig:SRM_P_max} compares the achievable sum rate of FIM-enabled multiuser communication with that of RA-enabled communication. 
The FIM-enabled NF system, utilizing CSI obtained from the proposed HSABT scheme, exhibits only a marginal performance gap compared with the ideal case of perfect CSI, yet achieves more than 10 bps/Hz higher sum rate than the RA-based counterpart. 
In addition, the digital beamformer optimized via the WMMSE algorithm demonstrates a clear advantage over the zero-forcing (ZF) beamformer, further improving the multiuser sum rate performance.
As shown in Fig.~\ref{fig:SRM_K}, the proposed FIM-enabled multiuser communication consistently outperforms the RA-enabled counterpart across different numbers of users, and the performance gain becomes more pronounced as the number of users increases. 
These results indicate that FIM not only plays a crucial role in improving NF BT accuracy but also significantly enhances the communication quality in NF multiuser systems. Hence, FIM is expected to be a key enabling technology for future intelligent communication networks.

\section{Conclusion}
In this paper, we investigated XL-FIM-enabled hybrid near-far-field communications under practical discrete PS constraints. To mitigate phase-quantization errors, we established the corresponding FIM channel and beam-pattern models and developed a surface-shaping-based compensation strategy. The deformation-range analysis showed that wavelength-scale shaping is sufficient to achieve beam patterns comparable to those obtained with continuous PSs. We further proposed a low-overhead HSABT framework with adaptive NF-FF codeword switching and multi-path candidate tracking, achieving performance comparable to or better than exhaustive polar-codebook training with substantially reduced overhead. For high-gain multiuser transmission, an optimization-based method and a non-iterative KOM were developed. Simulation results demonstrated considerable improvements over conventional rigid arrays in beam-pattern accuracy, beam-training performance, and multiuser sum rate. Overall, jointly exploiting electronic phase control and geometric surface shaping provides a promising solution for practical XL-array communications under stringent hardware constraints.


\bibliographystyle{IEEEtran}
\bibliography{refs}

\end{document}